\documentclass[11pt,a4paper,preprintnumbers]{article}

\usepackage{jheppub}
\usepackage{graphicx}
\usepackage{amsmath}
\usepackage{xspace}
\usepackage{xcolor}
\usepackage{float}
\usepackage{comment}
\usepackage{subcaption}
\usepackage{textcomp}
\usepackage{placeins}
\usepackage{bm}
\usepackage{pstricks}
\usepackage{color}
\usepackage{braket}
\usepackage{slashed}
\usepackage{soul}
\usepackage{graphicx}

\newcommand{\refcite}[1]{Ref.~\cite{#1}}

\newcommand{\Fig}[1]{Fig.~\ref{fig:#1}}

\newcommand{\tab}[1]{Tab.~\ref{tab:#1}}

\newcommand{\ord}[1]{\mathcal{O}(#1)}

\newcommand{\rescalethreeplots}{0.33\textwidth}

\newcommand{\as}{\alpha_{\rm s}}

\newcommand{\Tau}{\mathcal{T}}
\newcommand{\GeV}{\,\mathrm{GeV}}

\newcommand{\cut}{\mathrm{cut}}

\newcommand{\de}{\mathrm{d}}

\newcommand{\geneva}{\textsc{Geneva}\xspace}
\newcommand{\sherpa}{\textsc{Sherpa}\xspace}
\newcommand{\openloops}{\textsc{OpenLoops}}
\newcommand{\Matrix}{\textsc{Matrix}\xspace}
\newcommand{\pythiaEight}{\textsc{Pythia8}\xspace}
\newcommand{\hhgrid}{\textsc{HHgrid}\xspace}
\newcommand{\scetlib}{\textsc{SCETlib}\xspace}
\newcommand{\dire}{\textsc{Dire}\xspace}
\newcommand{\ftapprox}{FT$_{\mathrm{approx}}$\xspace}

\begin{document}
%%%%%%%%%%%%%%%%%%%%%%%%%%%%%%%%%%%%%%%%%%%%%%%%%%%%%%%%%%%%%%%%%%%%%%%%%%%%%%%%
% Title page
%%%%%%%%%%%%%%%%%%%%%%%%%%%%%%%%%%%%%%%%%%%%%%%%%%%%%%%%%%%%%%%%%%%%%%%%%%%%%%%%

\title{NNLO+PS Double Higgs boson production with top-quark mass corrections in GENEVA}
\preprint{\vbox{\hbox{DESY-25-074}\hbox{}}}

\author[a]{Simone Alioli,}

\author[b]{Giulia Marinelli,}

\author[a]{and Davide Napoletano}

\affiliation[a]{Universit\`{a} degli Studi di Milano-Bicocca \& INFN, Piazza della Scienza 3, Milano 20126, Italy\vspace{0.5ex}}
\affiliation[b]{Deutsches Elektronen-Synchrotron DESY, Notkestr. 85, 22607 Hamburg, Germany\vspace{0.5ex}}
\emailAdd{simone.alioli@unimib.it}
\emailAdd{giulia.marinelli@desy.de}
\emailAdd{davide.napoletano@unimib.it}

\date{\today}

%%%%%%%%%%%%%%%%%%%%%%%%%%%%%%%%%%%%%%%%%%%%%%%%%%%%%%%%%%%%%%%%%%%%%%%%%%%%%%%%
\abstract{
  We present the implementation of the NNLO QCD corrections to
  double Higgs boson production at hadron colliders in \geneva, matched to the
  parton shower. We include all the known top-quark mass effects and the
  resummation of large logarithms of the zero-jettiness $\Tau_0$, up to
  NNLL$^\prime$ accuracy.
  This work extends our previous study, which was performed in the $m_t\to
  \infty$ infinite top-quark mass approximation, providing a more realistic
  simulation framework for Higgs boson pair production.
  We validate our approach against NNLO predictions by \Matrix and assess the
  importance of mass effects comparing with our $m_t\to \infty$ previous
  implementation.
}
%%%%%%%%%%%%%%%%%%%%%%%%%%%%%%%%%%%%%%%%%%%%%%%%%%%%%%%%%%%%%%%%%%%%%%%%%%%%%%%%

\maketitle
\flushbottom

%%%%%%%%%%%%%%%%%%%%%%%%%%%%%%%%%%%%%%%%%%%%%%%%%%%%%%%%%%%%%%%%%%%%%%%%%%%%%%%%
\section{Introduction}
\label{sec:intro}
With the upcoming High-Luminosity phase of the Large Hadron Collider (LHC),
one of the most urgent tasks is to tighten the constraint of the
Higgs boson trilinear coupling.
While most of its properties have been precisely
measured and shown to be consistent with Standard Model predictions~\cite{CMS:2020xrn,ATLAS:2023owm,ATLAS:2015zhl},
only loose bounds currently exist on the Higgs boson self-coupling~\cite{ATLAS:2024ish}.

The most important process entering the study of the Higgs boson trilinear
coupling is the production of a pair of Higgs bosons, where the self-coupling
appears already at leading order~\cite{Glover:1987nx,Eboli:1987dy,Plehn:1996wb}.
Due to its similarity to the single Higgs boson production case, higher order
QCD corrections (NNLO~\cite{deFlorian:2013uza,deFlorian:2016uhr} and N$^3$LO~\cite{Chen:2019lzz})
have been computed in the infinite top-quark mass limit,
but NLO QCD corrections including the full top-quark mass dependence
are also available~\cite{Borowka:2016ehy,Borowka:2016ypz}.
In addition, electroweak (EW) corrections have been recently presented~\cite{Heinrich:2024dnz,Bonetti:2025vfd},
as well as methods to include approximate top-quark mass corrections in higher order
calculations~\cite{Grigo:2013rya,Grigo:2014jma,Grigo:2015dia,Frederix:2014hta,Maltoni:2014eza,Maierhofer:2013sha}.
Finally, NLO QCD corrections with full top-quark mass dependence have been matched to
parton showers~\cite{Heinrich:2017kxx,Heinrich:2019bkc,Jones:2017giv}.

In this work, we extend our previous result in
\geneva~\refcite{Alioli:2022dkj}, where we presented fully
differential Higgs boson pair events at NNLO accuracy in the strict
$m_t\to\infty$ limit, matched to parton showers.
The exact inclusion of
top-quark mass effects at NNLO is extremely challenging,
as the process features a heavy-quark loop already at leading order.
Despite this, partial three-loops amplitudes, which are part of the
full NNLO corrections, are starting to become
available~\cite{Davies:2024znp,Hu:2025aeo,Davies:2025ghl}.
Here, we include full top-quark mass corrections at NLO and approximate mass
corrections at NNLO, achieving what is usually referred to as
\ftapprox, as presented at fixed order in~\refcite{DeFlorian:2018eng,Grazzini:2018bsd}.
Compared to these results, our implementation includes the resummation of
zero-jettiness, $\Tau_0$, at NNLL$^\prime$ accuracy, as well as parton
shower matching.
It is important to notice, however, that in the \ftapprox the inclusion of the unknown mass
corrections in the virtual amplitudes
is done via a reweighting procedure which originates
from the case of the production of a single Higgs boson in gluon fusion,
where the leading order amplitude involves only a single diagram -- the top-quark triangle loop.
As such, there is no ambiguity on whether the reweighting of these
contributions is performed at amplitude, or squared amplitude level, as they yield the same results.
For di-Higgs production, this is no longer the case, as the process now comprise
a resonant production channel which resembles that of the single Higgs boson and a non-resonant heavy-quark box
diagram.
The relative contributions of these diagrams vary across different regions of phase space,
allowing for multiple, formally equivalent, choices for the reweighting, leading to different approximations.
See~\refcite{Alasfar:2023xpc,Cadamuro:2025car} and references therein for more details,
where various parts of the amplitudes are reweighted differently.
In our case, we chose to reweight squared matrix
elements, as this significantly simplifies the implementation task.

Regardless of the details of the reweighting procedure used for the contributions that are still unknown, the inclusion of top-quark mass effects
plays a crucial role in achieving realistic and
accurate predictions for the production of a pair of Higgs bosons.
Indeed, it is widely accepted that the $m_t \to \infty$ approximation
only works in a narrow region of phase space and is thus not suitable
for precise phenomenological applications.

In addition to higher-order QCD corrections, recent calculations have also
provided higher-order EW corrections~\cite{Bi:2023bnq,Heinrich:2024dnz,Bonetti:2025vfd},
which are essential for reducing theoretical uncertainties.
Since this process is proportional to the top-quark Yukawa coupling, it features a strong $m_t$
scheme dependence, resulting in an estimated uncertainty of
approximately $20\%$~\cite{Bagnaschi:2023rbx}.
This can only be reduced by
including higher-order top-quark Yukawa corrections, which are of EW nature.
In this work, however, we neglect such effects, as our primary interest is
on the effect of QCD corrections.
Nonetheless, future versions of our code will need to
incorporate top-quark Yukawa scheme variations, as well as at least approximate
EW corrections.
Another source of theoretical uncertainty arises from the treatment of bottom-quark mass
corrections and of possible interference terms.
To our current knowledge, the only estimates of these effects come from
single Higgs boson production, where they are known to be around
5\%~\cite{Grazzini:2013mca,Mueller:2015lrx}.
Two-loop corrections, including bottom-quark mass, which are part of
the NLO corrections, have only been recently computed in the planar limit~\cite{Hu:2025aeo}.
We therefore neglect bottom-quark mass effects in our calculation, treating the
bottom quark as massless throughout.

Since the production of a pair of Higgs bosons is highly sensitive to the Higgs boson self-coupling,
it offers the possibility to probe eventual differences with the Higgs boson potential
as predicted by the Standard Model.
For this purpose, this process has been heavily studied within the Standard Model Effective Theory (SMEFT)
framework, including NLO QCD corrections and parton
shower effects~\cite{Grober:2015cwa,Grober:2017gut,deFlorian:2017qfk,Alioli:2018ljm,Heinrich:2020ckp,Dawson:2021xei,deFlorian:2021azd,Heinrich:2022idm,Alasfar:2023xpc,Isidori:2023pyp,DiNoi:2023ygk,Heinrich:2023rsd,Heinrich:2024rtg}.

The outline of this paper is the following.
In Sec.~\ref{sec:theory}, we highlight the differences of this new implementation with
respect to~\refcite{Alioli:2022dkj}, describing, in particular, the details of the reweighting
procedure used to obtain the \ftapprox predictions.
In Sec.~\ref{sec:validation}, we describe the validation of our findings and compare with results from~\refcite{Grazzini:2018bsd}.
In Sec.~\ref{sec:nnlo_partonic}, we study the importance of the inclusion of top-quark mass effects by comparing
our parton level results obtained with different treatments of the top-quark mass corrections.
We then proceed to compare our best \ftapprox predictions at different
stages of the parton shower procedure  in Sec.~\ref{sec:showered-results}.
Finally, in Sec.~\ref{sec:conclusions}, we report our conclusions.
%%%%%%%%%%%%%%%%%%%%%%%%%%%%%%%%%%%%%%%%%%%%%%%%%%%%%%%%%%%%%%%%%%%%%%%%%%%%%%%%

%%%%%%%%%%%%%%%%%%%%%%%%%%%%%%%%%%%%%%%%%%%%%%%%%%%%%%%%%%%%%%%%%%%%%%%%%%%%%%%%
\section{Theoretical framework and differences with $m_t \to \infty$ approximation}
\label{sec:theory}
In this work, we extend our previously presented implementation of
Higgs boson pair production process in \geneva~\cite{Alioli:2022dkj} to
include top-quark mass corrections up to the highest currently known order.
Following the common convention in the literature, we refer to this result as
\ftapprox~\cite{Frederix:2014hta,Maltoni:2014eza,Grazzini:2018bsd}.
We start by briefly reviewing the main features of the existing implementation,
referring the reader to~\refcite{Alioli:2022dkj} and references therein for
a more
comprehensive discussion of the \geneva framework and its
implementation.
Here, we limit ourselves to report the main formulae that are either essential
for our discussion of mass effects, or are necessary to describe our modifications.
In addition to our new \ftapprox result, we also compare to our previous
$m_t \to \infty$ approximation
as well as to the leading order reweighted result (referred to as B-proj).
These results all share the same structure. The interested reader can find all the
relevant formulae for the differential cross sections at different multiplicities
in Eqs.~(2.2-2.4) of~\refcite{Alioli:2022dkj}.

Calculations in the \geneva framework are done using a resolution
variable to split the phase space into resolved and unresolved QCD emissions.
A common choice for this variable is the zero-jettiness
$\Tau_0$~\cite{Stewart:2010tn}, which we employ in this work.
In the region where the resolution variable is smaller than a given
$\Tau_0^{\mathrm{cut}} \ll Q$, we can
exploit the leading power SCET factorisation theorem for this observable~\cite{Stewart:2009yx}.
In particular, this allows us to perform the resummation of large logarithms of $\Tau_0^{\mathrm{cut}} /Q$,
where $Q = m_{HH}$ is the invariant mass of the Higgs boson pair system.
At leading power in $\Tau_0^{\mathrm{cut}} / m_{HH}$, we can write the differential cross section
as
\begin{equation}\label{eq:convolutionmess}
  \frac{\de \sigma^{\rm SCET}}{\de \Phi_0 \, \de \Tau_0} =
  H_{{\scriptscriptstyle gg \to HH }}(Q^2,\mu) \int B_g(t_a,x_a,\mu) \,
  B_g(t_b,x_b,\mu) \, S_{gg}\left( \Tau_0 - \frac{t_a+t_b}{Q}, \mu \right) \,
  \de t_a \, \de t_b,
\end{equation}
where $H$ stands for the hard function, while $S$ and $B$ for the soft and beam functions, respectively.
The resummed formula is then matched to the appropriate fixed order calculation to yield
predictions that are valid across the entire phase space.
This is achieved in an additive approach, by summing the resummed and fixed order contributions and
subtracting the appropriate fixed order expansion of the resummed contribution (resummed-expanded).
Note that Eq.~\eqref{eq:convolutionmess} is identical, in form, to Eq.~(2.17) of~\refcite{Alioli:2022dkj},
although the exact top-quark mass dependence of the perturbative ingredients within each term does depend on the
specific approximation.
In the following subsections we detail these differences in the three cases considered in this work.

\subsection{The \ftapprox approximation}
The \ftapprox is defined by including the exact mass effects up to NLO and, where possible, also at NNLO.
For NNLO contributions where the exact top-quark mass dependence is not known,
the unknown terms are reweighted using the corresponding Born-level contributions with exact
mass dependence~\cite{Frederix:2014hta,Maltoni:2014eza,Grazzini:2018bsd}.

All Born-level processes are computed with our existing
\openloops~\cite{Cascioli:2011va,Buccioni:2017yxi,Buccioni:2019sur} interface, which provides
the exact top-quark mass dependence.
These include the loop-induced leading order diagram, as well as all real and double-real corrections.
Exact virtual corrections are only known up to NLO and are
available through \hhgrid~\cite{Borowka:2016ehy,Borowka:2016ypz,Heinrich:2017kxx},
which provides pre-computed amplitudes on a two-dimensional grid for fixed Higgs
boson and top-quark masses.
These grids are then interpolated to avoid re-computation of the full two-loop
integrals at every phase space point, and we have implemented a dedicated
interface to access the value of these grids in \geneva.
NNLO double-virtual and real-virtual corrections are not known with full top-quark mass
dependence. These are therefore reweighted by the corresponding Born matrix elements computed
with exact mass dependence.
Compared to the original formulae presented in Eqs.~(2.2-2.4) of~\refcite{Alioli:2022dkj},
we compute the $B_0, B_1, B_2$ and $V_0$ terms using exact top-quark mass dependence.
We include leading order mass effects in the real-virtual term $V_1$, through the reweighting
\begin{equation}
  V_{1}\left(\Phi_1\right) = V_{1}\left(\Phi_1,m_t\to\infty\right)
  \frac{B_1\left(\Phi_1,m_t\right)}
  {B_1\left(\Phi_1,m_t\to\infty\right)}\,.
\end{equation}
As explained in Eq.~(2.27) of~\refcite{Alioli:2022dkj}, we do not need to
reweight the double-virtual
term $W_0$, as it does not contribute for $\Tau_0>\Tau_0^{\mathrm{cut}}$.
Instead, two-loop virtual corrections are included in the two-loop hard function coefficient
via Eq.~\eqref{eq:convolutionmess}.
The only change in the resummation part of our calculation, compared to the $m_t \to \infty$ result,
lies in the top-quark mass dependence of the hard function perturbative coefficients,
which encode the hard part of loop corrections.
This is a consequence of the universality of the factorisation theorem:
since soft and collinear modes are insensitive to the details of the hard modes that are integrated out,
top-quark mass effects enter exclusively through the Wilson coefficient of the hard function.
In this case, the Born and the finite part of the one-loop coefficients,
which are related to $H^{(0)}$ and $H^{(1)}_{\mathrm{fin}}$, respectively,
are known exactly and obtained from
\openloops~\cite{Cascioli:2011va,Buccioni:2017yxi,Buccioni:2019sur} and
\hhgrid~\cite{Borowka:2016ehy,Borowka:2016ypz,Heinrich:2017kxx}.
However, to ensure cancellation of the slicing variable dependence,
the fixed order expansion of the resummed expression must be consistent with
the fixed order implementation.
In particular, for
$\Tau_0>\Tau_0^{\mathrm{cut}}$, the finite one-loop contribution $H^{(1)}_{\mathrm{fin}}$
must be reweighted in the same way as the real-virtual term.
Accordingly, the two-loop term, which appears in the finite part of the second order
coefficients, $H^{(2)}_{\mathrm{fin}}$, and $H^{(1)}_{\mathrm{fin}}$
are thus rescaled as
\begin{equation}
  \label{eq:twoloophard}
  H^{(i)}_{\mathrm{fin}}(\Phi_0) = H^{(i)}_{\mathrm{fin}}\left(\Phi_0,m_t\to\infty\right)
  \frac{B_0\left(\Phi_0,m_t\right)}
  {B_0\left(\Phi_0,m_t\to\infty\right)}\,, \quad i=1,2\,.
\end{equation}
We remark that the reweighting of the top-quark mass
dependence in $H^{(1)}_{\mathrm{fin}}$ is only necessary to ensure the
correct expansion of the resummed cross section and its cancellation
against the $V_1$ contribution, which is reweighted similarly. This
does not mean that the exact top-quark mass dependence at NLO coming
from \hhgrid is not used in our calculation, because it is still
present in the fixed order part, where it enters in the virtual
correction, $V_0$.  Alternatively, one could have also included the
exact $H^{(1)}_{\mathrm{fin}}$ in the resummed part and the
approximated one in the resummed-expanded. In this way one would get the exact
top-quark mass dependence at NNLL$^\prime$ but then
the cancellation between the resummed and the resummed-expanded at large values of $\Tau_0$
would need to be enforced by hand.
In the current implementation we preferred the first option, \emph{i.e.}~to include
the exact top-quark mass dependence only at fixed order.

The relation between the Born, the
finite part of the one-loop amplitudes, and the hard function
coefficients can be found \emph{e.g.}~in~\refcite{Alioli:2022dkj,Becher:2009qa}.
All coefficients are then
transformed to the $\overline{\rm MS}$
scheme~\cite{Becher:2009qa,deFlorian:2012za} from the IR subtraction
scheme in which they were originally computed. Crucially, all other
terms, either in the soft and beam functions or in the evolution
operators needed for the $\Tau_0$ and $\Tau_1$ resummations, remain
unchanged from the $m_t\to\infty$ case.  This includes our choice of
profile scales and the relevant transition points,
see~\refcite{Alioli:2022dkj} and references therein for more details.

Given that the loop-order of almost all matrix elements is increased by
one unit in the \ftapprox, the calculation is significantly more
challenging from the numerical point of view.  In particular, the
treatment of the one-loop double-real matrix element $gg \to HH gg$ is
notably delicate near the double limit $\Tau_0 , \Tau_1 \to 0$,
compared to the other approximations. The stability of double-real matrix
elements near the double unresolved limit is an issue that has been noted also
in~\refcite{Grazzini:2018bsd}.
Their approach was to use the $m_t \to \infty$ matrix element for $gg \to HHgg$,
reweighted by the massive Born matrix element in regions where at least one scalar
product $\alpha_{ij} = p_i\cdot p_j/\hat{s}$ falls below a technical cut,
chosen to be  $\alpha_{\mathrm{cut}} = 10^{-4}$.
The mapping used to project the two-parton phase space into the $gg\to HH$ kinematics
is done by preserving the momenta of the two Higgs bosons, but significantly affects the incoming partons,
which might no longer be along the beam directions.
In our case, such procedure cannot be used, as it breaks the subtraction
mechanism required for the calculation of the NLO corrections to the one-parton
phase space (NLO$_1$).
We therefore construct the approximation
of the double-real matrix elements as follows
\begin{equation}
  {\de\sigma_{gg\to HH
      gg}\left(\Phi_2\right)\Biggr|}_{\min(\alpha_{ij})<\alpha_{\mathrm{cut}}}
  = {\de\sigma_{gg\to HH g}\left(\widetilde{\Phi}_1\right)}
  \frac{\de\sigma_{gg\to HHgg}^{m_t\to\infty}\left(\Phi_2\right)}
  {\de\sigma_{gg\to HH g}^{m_t\to\infty}\left(\widetilde{\Phi}_1\right)}
\end{equation}
where $\widetilde{\Phi}_1$ is obtained by applying the
FKS projection mapping~\cite{Frixione:1995ms} to the original
two-parton phase space point $\Phi_2$.
Similarly to \refcite{Grazzini:2018bsd}, we adopt $\alpha_{\mathrm{cut}} = 10^{-4}$ as our technical cutoff.
The rationale behind this reweighting choice is that, in
the soft-collinear limit, the ratio ${\de\sigma_{gg\to HHgg}^{m_t\to\infty}\left(\Phi_2\right)}
/{\de\sigma_{gg\to HH g}^{m_t\to\infty}\left(\widetilde{\Phi}_1\right)}$ tends towards
the splitting function used in the subtraction, thus reproducing the structure of
the subtraction term. This ensures that the cancellation between the reweighted $gg\to HHgg$ matrix element and its counterterms, which are always evaluated with the exact top-quark mass dependence, are always locally finite in each singular limit when $\alpha_{\mathrm{cut}} \to 0$.
Note that, however, for any given finite value of $\alpha_{\mathrm{cut}}$, the projected phase space point $\widetilde{\Phi}_1$ does
not exactly match the point $\Phi_1$ at which the subtraction term is
evaluated. Their difference, due to IR safety, is
suppressed by powers $\mathcal{O}(\alpha^n_{\cut})$. Consequently, in regions where
$\mathcal{O}(\alpha_{\cut})\sim\mathcal{O}(\Tau_0^{\cut}/Q)$, we expect to observe
numerical discrepancies as well as instabilities in the subtractions procedure.
To mitigate this, given that the smallest possible value of $\alpha_{\mathrm{cut}}$ is dictated by the stability of
the $gg \to HH gg$ matrix element, the value of $\Tau_0^{\cut}$ must be chosen to be much larger than $\alpha_{\mathrm{cut}} Q$.
A more detailed discussion and validation of this
choice is presented in Sec.~\ref{sec:validation}.

\subsection{Comparison to other approximations}
We compare our \ftapprox results with other approximations existing in the literature,
namely the $m_t\to\infty$ approximation, corresponding to our previous results of~\refcite{Alioli:2022dkj},
and the B-proj approximation as described in~\refcite{Grazzini:2018bsd}.
Below, we summarize the main features of these two approximations and refer the reader to the
corresponding references for further details.

\paragraph{\textbf{\underline{\boldmath $m_t \to \infty$}}:} This corresponds to the exact infinite top-quark mass limit,
as described in~\refcite{Alioli:2022dkj}, \textit{i.e.}~with no rescaling or mass reweighting applied.
In the literature, this approximation is often referred to as HTL, or HEFT, but in this work we reference it with
$m_t\to\infty$ to avoid confusion.

\paragraph{\underline{B-proj}:} The B-proj approximation consists of reweighting all squared matrix elements
-- including those entering the hard function perturbative coefficients -- using Born
projected matrix elements. Specifically, the reweighting factor
is given by
\begin{equation}
  \label{eq:bprojkfact}
  K(\Phi_n) = \frac{B_{gg\to HH}\left(\tilde{\Phi}_0(\Phi_n),m_t\right)}
  {B_{gg\to HH}\left(\tilde{\Phi}_0(\Phi_n),m_t\to\infty\right)}\,.
\end{equation}
Notice that this provides the exact top-quark mass dependence only at
LO. Starting from NLO and in the presence of further emissions, the
reweighting factor is calculated on a projected kinematic
$\tilde{\Phi}_0(\Phi_n)$, taken from Eqs.~(2.2) and~(2.3) of~\refcite{Grazzini:2018bsd},
and defined in~\refcite{Catani:2015vma}.
%%%%%%%%%%%%%%%%%%%%%%%%%%%%%%%%%%%%%%%%%%%%%%%%%%%%%%%%%%%%%%%%%%%%%%%%%%%%%%%%

%%%%%%%%%%%%%%%%%%%%%%%%%%%%%%%%%%%%%%%%%%%%%%%%%%%%%%%%%%%%%%%%%%%%%%%%%%%%%%%%
\section{Validation of the NNLO result}
\label{sec:validation}
% -------------------------------------------------------------------------------
\begin{figure}[htb]
\begin{center}
  \includegraphics[width=0.6\textwidth]{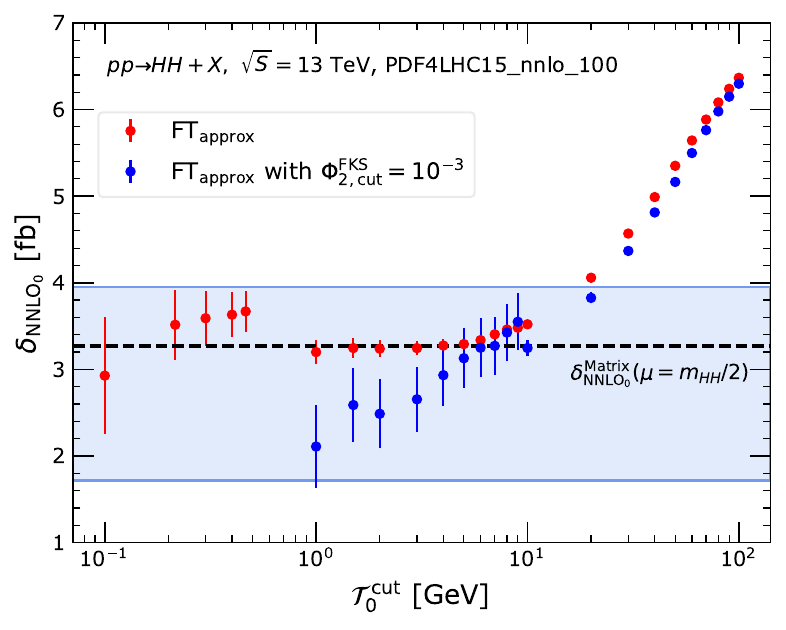}
  \caption{
  Relative NNLO corrections as a function of the slicing parameter $\mathcal{T}_0^{\cut}$,
  evaluated at $\sqrt{S} = 13$ TeV and $\mu = m_{HH}/2$, using the \texttt{PDF4LHC15\_nnlo\_100}
  PDF set.
  The expected NNLO contribution from \Matrix, corresponding to 3.27 fb, is shown as a dashed
  black line.
  }\label{fig:tau0cut_slicing}
\end{center}
\end{figure}
%-------------------------------------------------------------------------------
We validate our results by comparing them to those in
~\refcite{Grazzini:2018bsd}, obtained using the \Matrix framework.
As discussed in detail in~\refcite{Alioli:2022dkj}, the partonic predictions from \geneva
cannot exactly match those from purely fixed order generators such as \Matrix.
This is because \geneva includes higher order effects coming from the
resummation of logarithms of the resolution variable and consequent matching to
fixed order.
Additionally, as discussed in Sec.~\ref{sec:theory},
we employ a different rescaling below the
$\alpha_{\mathrm{cut}}$ to mitigate instabilities in matrix elements evaluation
in the deep IR limit.
Therefore, in order to validate our predictions meaningfully, we calculate the pure NNLO correction
to the cross section as done in a pure slicing calculation. This means that for each value of the
resolution variable $\Tau_0^{\mathrm{cut}}$, we compute two contributions
separately. Below this value, the cross section is approximated by the cumulative
distribution of the resummation formula expanded at fixed order
$\mathcal{O}(\alpha_s^2)$. Above the cut, we compute the exact fixed order
contribution, which corresponds to the NLO correction to the $pp\to HH + j$
process, dubbed NLO$_1$. Schematically
\begin{equation}
  \label{eq:slice}
  \delta_{\mathrm{NNLO}_0} =
  \Sigma_{\mathcal{O}(\alpha_s^4)}\left(\Tau_0 < \Tau_0^{\cut}\right)
  +  \delta_{\mathrm{NLO}_1}\left(\Tau_0 > \Tau_0^{\cut}\right)
  + \mathcal{O}\left(\Tau_0^{\cut}\right)
\,.\end{equation}
The residual $\Tau_0^{\cut}$ dependence vanishes up to power corrections,
since the contribution below the cut is accurate only at leading power in
$\Tau_0$.

Figure~\ref{fig:tau0cut_slicing} shows the
reconstructed $\delta_{\rm NNLO_0}$ using the \ftapprox approximation
at $\sqrt{S} = 13$ TeV and $\mu = m_{HH}/2$, with the
\texttt{PDF4LHC15\_nnlo\_100} PDF set~\cite{Butterworth:2015oua}.  Red
dots represent the result of Eq.~\eqref{eq:slice}, compared to the
reference dashed black line at 3.27~fb, obtained by subtracting the
NNLO$_{\rm FT_{approx}}$ and the NLO \Matrix results from Table 1
in~\refcite{Grazzini:2018bsd}.  The light blue band around that line
corresponds to the scale variation values quoted in the same
reference.  We find good agreement between our predictions
and those obtained by \Matrix for small
($\mathcal{O}(1)$~GeV)\footnote{ This is in agreement with what
  explained in Sec.~\ref{sec:theory} since the average value of $Q$,
  which in our case corresponds to the invariant mass of the Higgs
  boson pair system, is around 500~GeV, so we expect to see numerical
  effects, due to a finite value of $\alpha_{\mathrm{cut}}$, until around
  $\Tau_0^{\cut} \lesssim 0.5$~GeV.}
values of
$\Tau_0^{\cut}$, where power corrections are suppressed.  As the
slicing parameter increases, deviations become visible as expected due
to the growing importance of neglected higher-order terms.  This
reinforces the choice of a small $\Tau_0^{\cut}$ for reliable and
efficient resummed predictions. Note that for smaller values of
$\Tau_0^{\cut}$ we see deviations and even numerical instabilities, as
discussed in Sec.~\ref{sec:theory}.  To show the impact of this
approximation, in Fig.~\ref{fig:tau0cut_slicing} we additionally
include, with blue dots, the effect of keeping the full top-quark mass
dependence without doing any type of approximation. Due to the matrix elements instability
problems mentioned earlier, however, the only way of
producing any meaningful result when using the exact $gg\to HH gg$
matrix element with full top-quark mass dependence is by using a cut
on the FKS variables $\xi$ and $y$ of the extra radiation phase
space. This cut is chosen to be $10^{-3}$, which is the lowest
possible value we can get to before the matrix elements become completely unstable.
We dub this
result ``FT$_{\mathrm{approx}}$ with
$\Phi_{2,\cut}^{\mathrm{FKS}}=10^{-3}$''.
Notice that the result obtained in this case is
only stable up to $\Tau_0^{\cut}\sim 10$~GeV, which would imply having
the result afflicted by larger fiducial power corrections.
This further validates our choice of approximating
the top-quark mass dependence of  $gg\to HH gg$ below the $\alpha_\cut$
value instead of keeping the exact mass dependence but be forced
to raise the technical cutoff: in this way we are lowering the  power correction in zero-jettiness
at the expense of including some power corrections to the exact top-quark mass.

% -------------------------------------------------------------------------------
\begin{figure}[htb]
  \begin{center}
    \includegraphics[width=\rescalethreeplots]{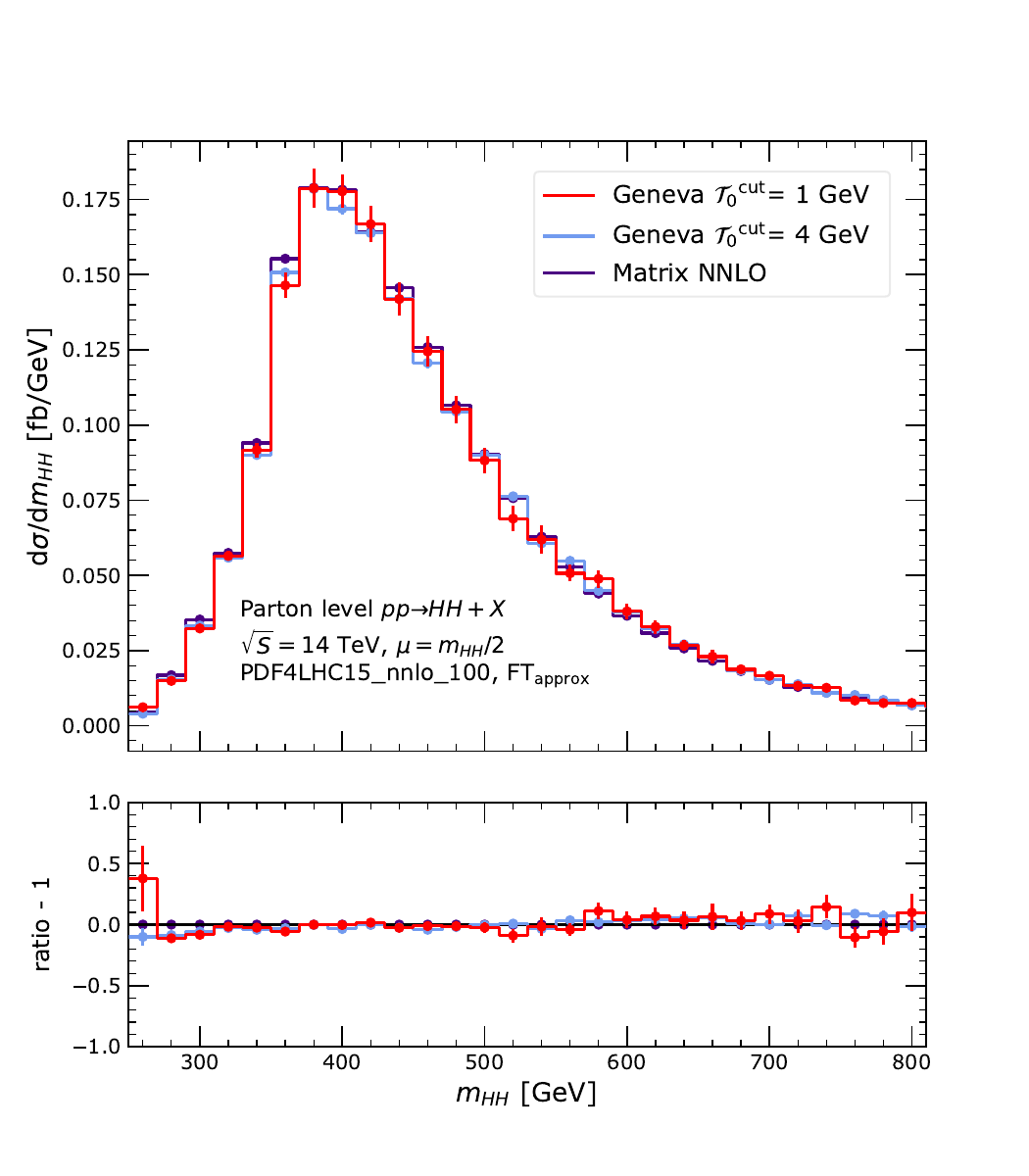}\hfill
    \includegraphics[width=\rescalethreeplots]{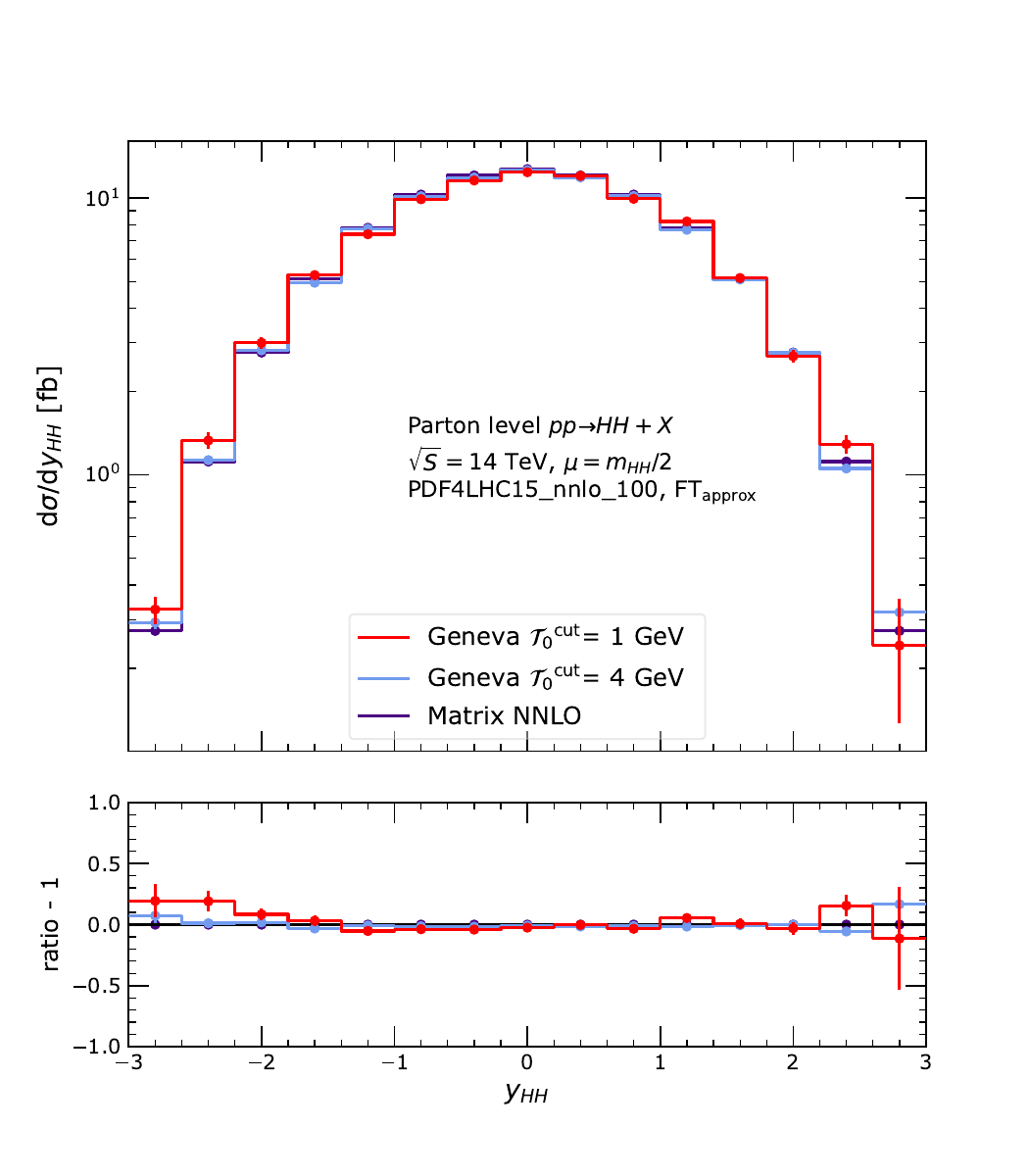}\hfill
    \includegraphics[width=\rescalethreeplots]{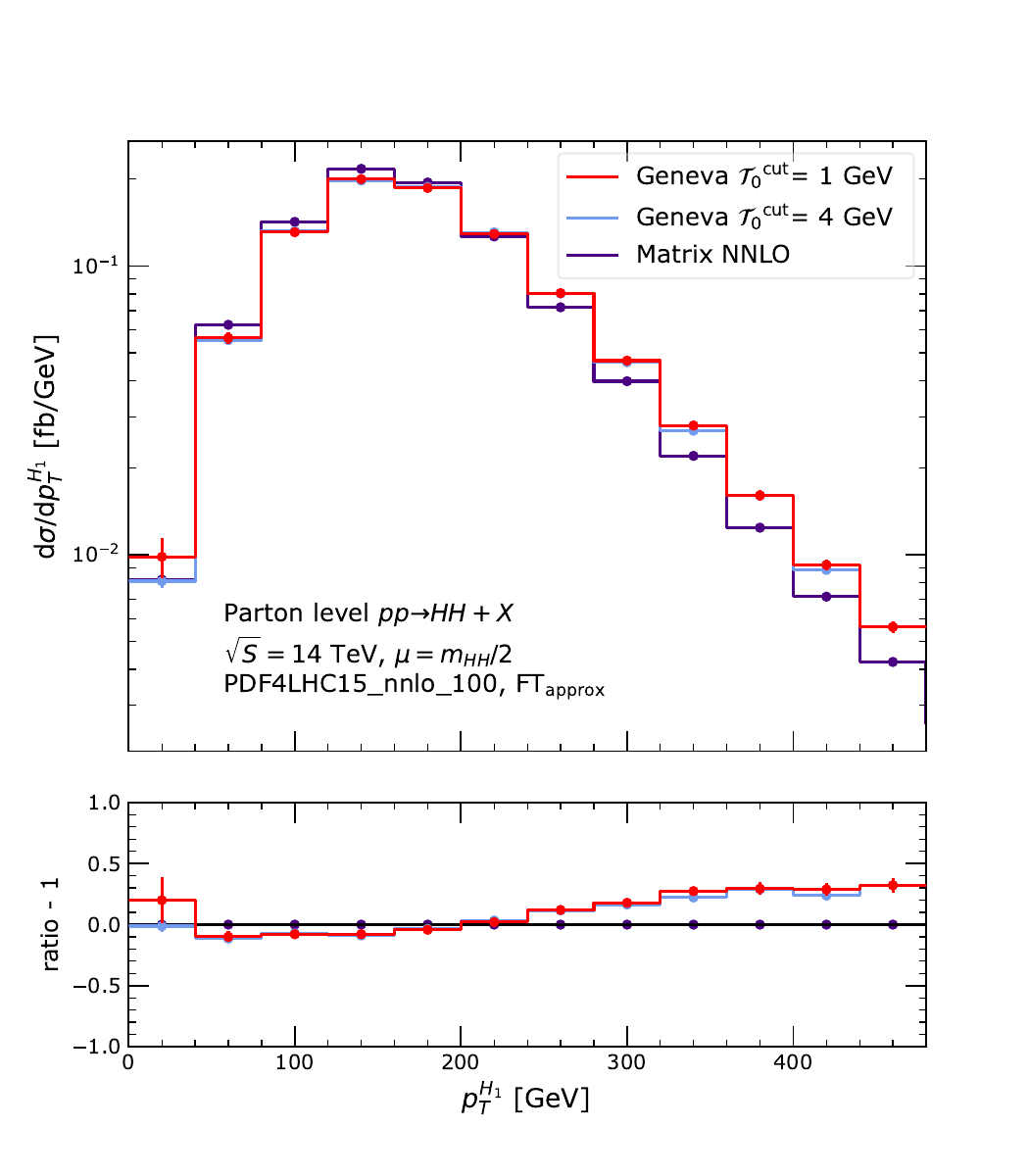}
  \end{center}
  \caption{
  Comparison between the full \geneva result for different
  $\Tau_0^{\mathrm{cut}}$ values (1~GeV in red and 4~GeV in light blue) and the \Matrix
  prediction (in purple) in the \ftapprox approximation, for the invariant mass of the Higgs boson pair (left),
  their rapidity (centre) and the transverse momentum of the hardest Higgs boson (right).
  The ratio panel shows the relative differences between the predictions.
  Results are shown at $\sqrt{S} = 14$ TeV, $\mu = m_{HH}/2$ and using the
  \texttt{PDF4LHC15\_nnlo\_100} PDF set.
  }
  \label{fig:matrix_comparison}
\end{figure}
%-------------------------------------------------------------------------------

Having validated the NNLO accuracy of our implementation, we can now move to the study of the \geneva predictions.
In Fig.~\ref{fig:matrix_comparison} we compare with \Matrix our
NNLL$^\prime +$ NNLO prediction for a
set of differential observables at NNLO accuracy. In particular we show the
invariant mass of the Higgs boson pair ($m_{HH}$),
their rapidity ($y_{HH}$) and the transverse momentum of the hardest Higgs boson ($p_T^{H_1}$).
Statistical uncertainties for the \Matrix predictions were not provided,
which is why they are not
displayed in the main panels. Since the available results for \Matrix were
obtained at $\sqrt{S} = 14$~TeV, the \geneva results presented in the remainder
of this section are obtained with the same setup.
We observe excellent agreement between the \geneva and \Matrix results, with
any discrepancies coming from the effects of the inclusion of higher order
effects in \geneva. These can play a particularly important role in the $p_T^{H_1}$ distribution,
as thoroughly discussed in~\refcite{Alioli:2022dkj}.
In addition, we show how varying our choice of $\Tau_0^{\cut}$ affects
differential distributions, and we see no significant impact when going from
$\Tau_0^{\cut} = 1$~GeV to $\Tau_0^{\cut}=4$~GeV. Note that, in the following, we
always employ $\Tau_0^{\cut} = 1$~GeV for our predictions, unless otherwise specified.

%-------------------------------------------------------------------------------
\begin{figure}[htb]
  \begin{center}
    \includegraphics[width=0.47\textwidth]{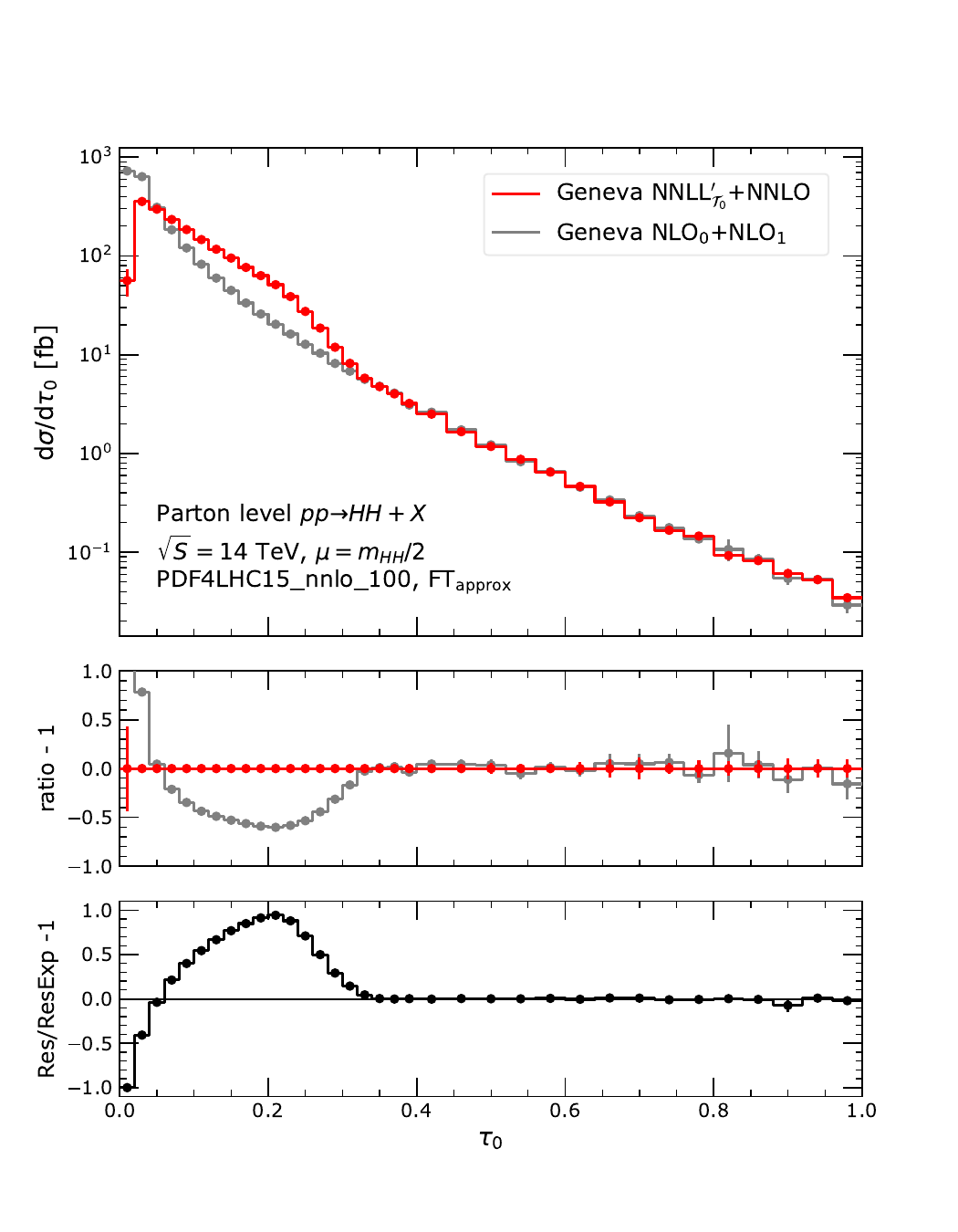}\hfill
    \includegraphics[width=0.47\textwidth]{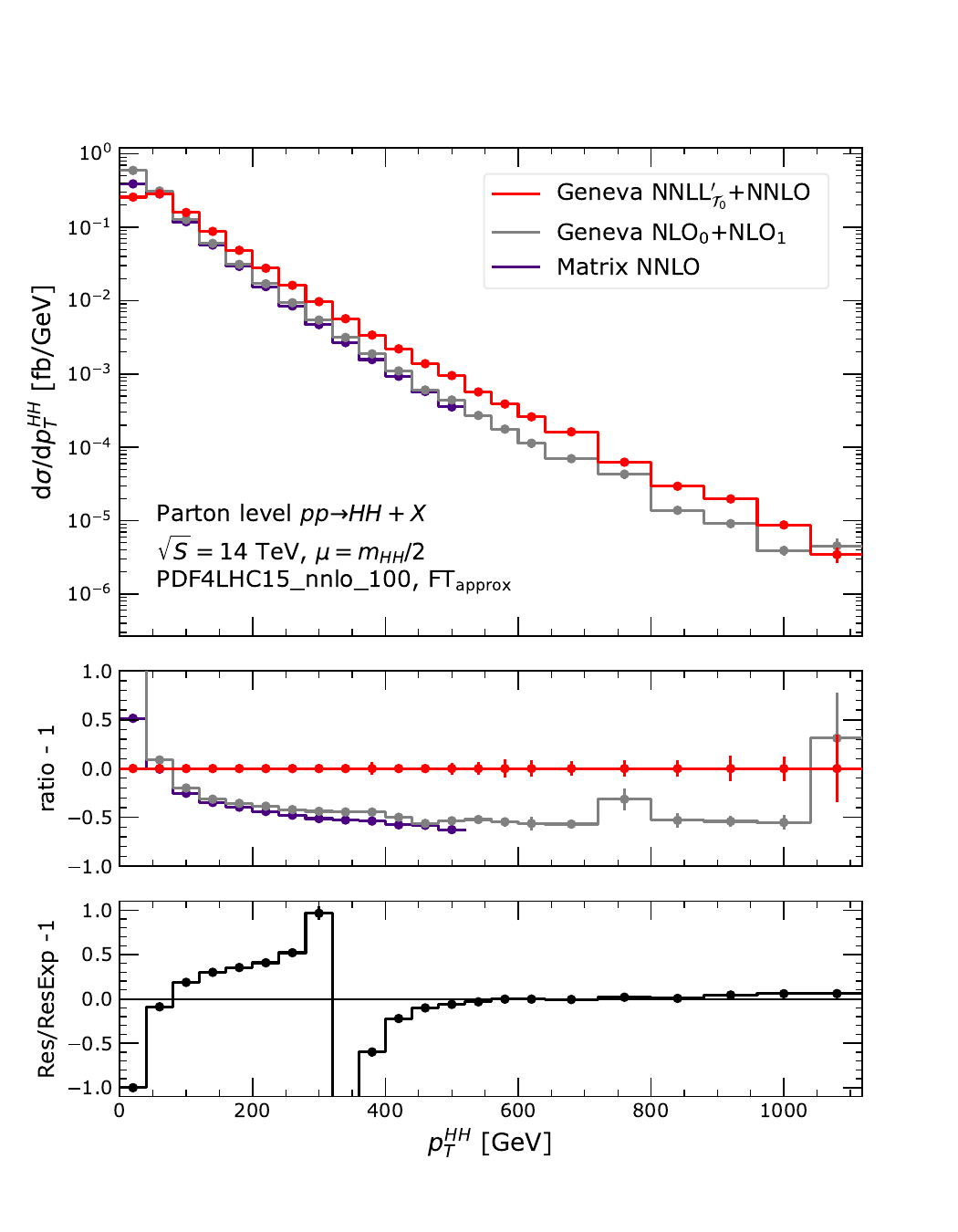}
  \end{center}
  \caption{
  Impact of the resummation on the reduced $\tau_0$ (left) and on the Higgs boson pair
  transverse momentum $p_T^{HH}$ (right).
  The full \geneva resummed result (in red) is compared with the corresponding fixed order prediction (in gray).
  For $p_T^{HH}$, we also report the \Matrix prediction (in purple).
  The top ratio panels show the relative differences between the predictions, while
  the lower panels show the contribution from the resummed part (i.e. the
  difference between the full resummed and its fixed order expansion).
  Results are shown at $\sqrt{S} = 14$ TeV, $\mu = m_{HH}/2$ and using the
  \texttt{PDF4LHC15\_nnlo\_100} PDF set.
  }
  \label{fig:resummation_impact}
\end{figure}
% -------------------------------------------------------------------------------
To further evidence the impact of resummation, we consider the \geneva result at
two levels: the first is NNLL$^\prime +$ NNLO, corresponding to our full partonic prediction,
while the second is the fixed order only contribution, obtained by running \geneva while switching
off both the resummed and the resummed-expanded contributions. Note that, strictly speaking,
this latter contribution -- shown in the plots for reference to illustrate the effect of
resummation in the hard region -- includes the $\ord{\as^2}$ terms only for $\Tau_0 > \Tau_0^{\cut}$.
We thus refer to it as NLO$_0$ +NLO$_1$.
We report both contributions in~\Fig{resummation_impact}
for the reduced $\tau_0 = \Tau_0/m_{HH}$ (left) and the Higgs boson pair transverse momentum
$p_T^{HH}$ (right).
For $\tau_0$, significant differences appear in the small-$\tau_0$ region, where
logarithmic corrections dominate and the resummation is crucial.
In the lowest panel, we report the relative difference between the resummation
and the resummed-expanded.
As it can be seen by comparing the first and the second sub-panels,
the difference between the matched result and the fixed-order
only component at small $\tau_0$ is entirely captured by the resummed term
alone, since it corresponds to the difference between the resummed and the
resummed-expanded. At larger values of $\tau_0$ the resummed and the
resummed-expanded cancel, as they should, since we expect the fixed order
expansion to be the correct approximation in this regime.

Crucially, the fact that we perform the resummation in $\tau_0$ has a significant
impact on the $p_T^{HH}$ distribution.
Since we switch off the resummation directly in $\tau_0$ instead of in $p_T^{HH}$,
the effects of the resummation extend at moderately large value of $p_T^{HH}$,
mainly because of the large size of $m_{HH}$.
This can be clearly seen in the right panel of~\Fig{resummation_impact}, where our
matched result shows large differences with respect to the \Matrix sample, which
are however entirely due to the presence of higher order effects from the
resummation.
Indeed, while we could only obtain \Matrix data for $p_T$ values up to $500$~GeV, we see
that at large $p_T^{HH}\gtrsim 1000~$GeV values, the full result and the
fixed order nicely converge, as they should.
A possible way to better control this behaviour would be to switch off the
resummation with a hybrid scale involving both $p_T$ and
$\Tau_0$~\cite{Cal:2023mib}, or perform a joint $p_T-\Tau_0$
resummation~\cite{Lustermans:2019plv}.
However, this goes beyond the scope of the present study and is left for future
investigation.
%-------------------------------------------------------------------------------
\begin{table}[ht]
\centering
\begin{tabular}{c|c|c|c}
\hline\hline
  \multicolumn{1}{c|}{cross sections [fb]}  & \multicolumn{1}{c|}{$\sqrt{S} = 13$ TeV} & \multicolumn{1}{c|}{$\sqrt{S} = 13.6$ TeV} & \multicolumn{1}{c}{$\sqrt{S} = 27$ TeV} \\
  \hline
  NNLO$_{\rm FT_{approx}}$ & $(31.19 \pm  0.38)^{+2.8\%}_{-5.7\%}$ & $(35.17 \pm 0.41)^{+2.7\%}_{-5.5\%}$ & $(134.44 \pm  3.47)^{+0.0\%}_{-2.41\%}$\\
 \hline\hline
\end{tabular}
\caption{
\geneva \ftapprox inclusive cross sections for Higgs boson pair production at $\sqrt{S} = 13, 13.6$ and $27$ TeV, computed at $\mu = m_{HH}/2$
The first uncertainty is statistical; the subscript/superscript indicate the
3-point inclusive scale variations. The PDF set used is
\texttt{PDF4LHC15\_nnlo\_100}.
}
\label{tab:xs_values_mhho2}
\end{table}
%-------------------------------------------------------------------------------

%-------------------------------------------------------------------------------
\begin{table}[ht]
\centering
\begin{tabular}{c|c|c|c}
\hline\hline
  \multicolumn{1}{c|}{cross sections [fb]}  & \multicolumn{1}{c|}{$\sqrt{S} = 13$ TeV} & \multicolumn{1}{c|}{$\sqrt{S} = 13.6$ TeV} & \multicolumn{1}{c}{$\sqrt{S} = 27$ TeV}\\
\hline
 NNLO$_{m_t \to \infty}$  & $(26.99 \pm 0.02)^{+7.2\%}_{-8.6\%}$ & $(30.69 \pm 0.02)^{+7.1\%}_{-8.5\%}$  & $(197.5 \pm 0.14)^{+6.8\%}_{-7.7\%}$\\
 NNLO$_{\rm B-proj}$      & $(33.79 \pm 0.03)^{+7.2\%}_{-8.4\%}$ & $(37.65 \pm 0.03)^{+7.3\%}_{-8.4\%}$  & $(166.8 \pm 0.12)^{+6.7\%}_{-7.4\%}$\\
 NNLO$_{\rm FT_{approx}}$ & $(29.11 \pm 0.20)^{+6.8\%}_{-5.2\%}$ & $(32.34 \pm 0.21)^{+6.8\%}_{-5.3\%}$ & $(132.9 \pm 1.20)^{+4.9\%}_{-3.4\%}$\\
 \hline\hline
\end{tabular}
\caption{
\geneva inclusive cross sections for Higgs boson pair production at $\sqrt{S} = 13, 13.6$ and $27$ TeV, computed at $\mu = m_{HH}$.
Results are presented for the three approximations: $m_t \to \infty$, B-proj and \ftapprox.
The first uncertainty is statistical; the subscript/superscript indicate the 3-point inclusive scale variations.
The PDF set used is \texttt{PDF4LHC21\_nnlo}.
}
\label{tab:xs_values_mhh}
\end{table}
%-------------------------------------------------------------------------------
To conclude the section, we report in~\tab{xs_values_mhho2} and~\tab{xs_values_mhh} the NNLO
\geneva inclusive cross sections (in fb) for the various top-quark mass approximations.
The predictions are computed at $\mu = m_{HH}/2$ and $\mu = m_{HH}$,
using the \texttt{PDF4LHC15\_nnlo\_100} and \texttt{PDF4LHC21\_nnlo} PDF
set~\cite{PDF4LHCWorkingGroup:2022cjn}, respectively.

We observe that the cross sections grow with $\sqrt{S}$, as expected from the increase
of parton luminosities, in particular the gluon luminosity. This is expected because  by increasing the
centre-of-mass energy we are probing the PDFs at relatively smaller values of
the parton longitudinal momentum fraction $x$.
Concerning the inclusion of the top-quark mass corrections, we see that the B-proj
approximation tends to overestimate the cross section compared to
\ftapprox, and we notice that its agreement with the \ftapprox result for total
cross sections is of similar size as that of the $m_t\to\infty$ approximation.
We also observe that scale uncertainties (obtained here by 3-point correlated $\mu_R,\mu_F$ variations) have a typical size of
$\sim 10\%$ at $\mu = m_{HH}$, consistent with previous $m_t \to \infty$
studies~\cite{Alioli:2022dkj}.
These uncertainties are reduced when running at $\mu = m_{HH}/2$, aligning better
with~\refcite{Grazzini:2018bsd}.
Future work will include 7-point scale variations for a more comprehensive uncertainty
estimate.
%%%%%%%%%%%%%%%%%%%%%%%%%%%%%%%%%%%%%%%%%%%%%%%%%%%%%%%%%%%%%%%%%%%%%%%%%%%%%%%%

%%%%%%%%%%%%%%%%%%%%%%%%%%%%%%%%%%%%%%%%%%%%%%%%%%%%%%%%%%%%%%%%%%%%%%%%%%%%%%%%
\section{NNLO partonic results}
\label{sec:nnlo_partonic}
In this section, we study the effects of the inclusion of top-quark mass
corrections at the partonic level.
In the \geneva framework, this corresponds to a NNLL$_{\Tau_0}^\prime$ + NNLO
calculation, before the inclusion of parton shower effects.
We always consider the process $pp\to HH + X$, where the leading order contribution comes
solely from the gluon fusion production channel, and we require two on-shell Higgs
bosons in the final state.
We set the centre-of-mass energy to $\sqrt{S} = 13.6$~TeV to be compatible with
the Run~3 of the LHC.
Both the factorization and renormalization scales are dynamically chosen and set equal to the invariant mass of
the Higgs boson pair, $\mu_F = \mu_R = m_{HH}$.
We use the \texttt{PDF4LHC21\_nnlo} parton distribution function set~\cite{PDF4LHCWorkingGroup:2022cjn},
accessed via \textsc{LHAPDF6}~\cite{Buckley:2014ana}, including the corresponding value of the strong coupling
constant $\alpha_s(m_Z)$. The running of such coupling is performed at three-loop order by default.

We compare the three approximations presented in Sec.~\ref{sec:theory}, namely
the FT$_{\mathrm{approx}}$, B-proj and $m_t\to\infty$ calculations.
We set the following input parameters
\begin{equation}
  m_H = 125 \; \GeV, \qquad   G_F = 1.1663787 {\rm e}^{-5} \; \GeV^{-2}, \qquad m_t = 173 \; \GeV
\end{equation}
to ensure consistency with the grids provided by \hhgrid for the calculation of
virtual corrections in FT$_{\mathrm{approx}}$.
We evaluate the beam functions, needed for the resummed part of our
calculation, through the \texttt{beamfunc} module of \scetlib~\cite{Billis:2019vxg,scetlib}.
Finally, we set our resolution cutoff as $\Tau_0^{\cut} = \Tau_1^{\cut} = 1 \; \GeV$.

% -------------------------------------------------------------------------------
\begin{figure}[htb!]
\begin{center}
  \includegraphics[width=0.49\textwidth]{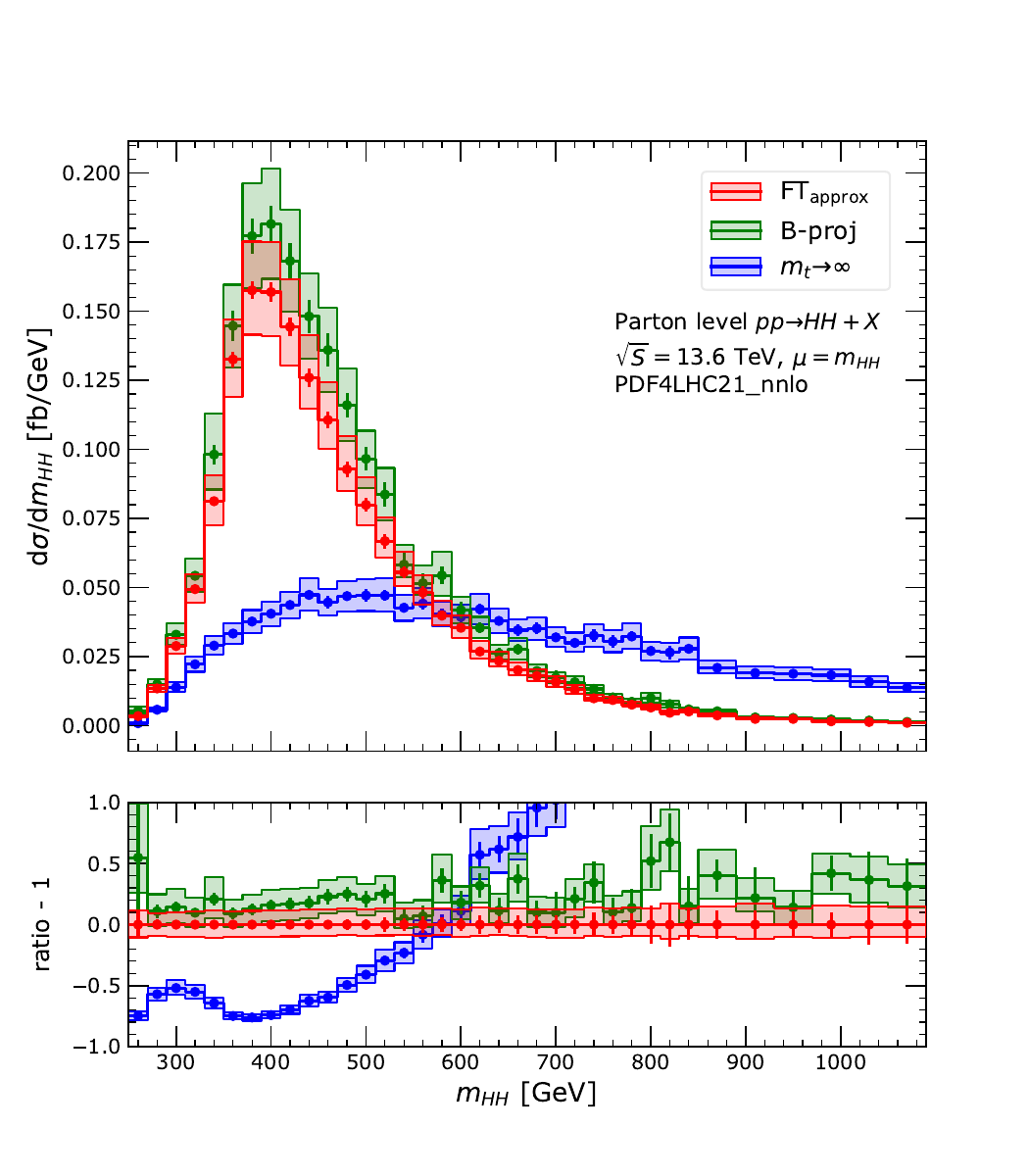} \hfill
  \includegraphics[width=0.49\textwidth]{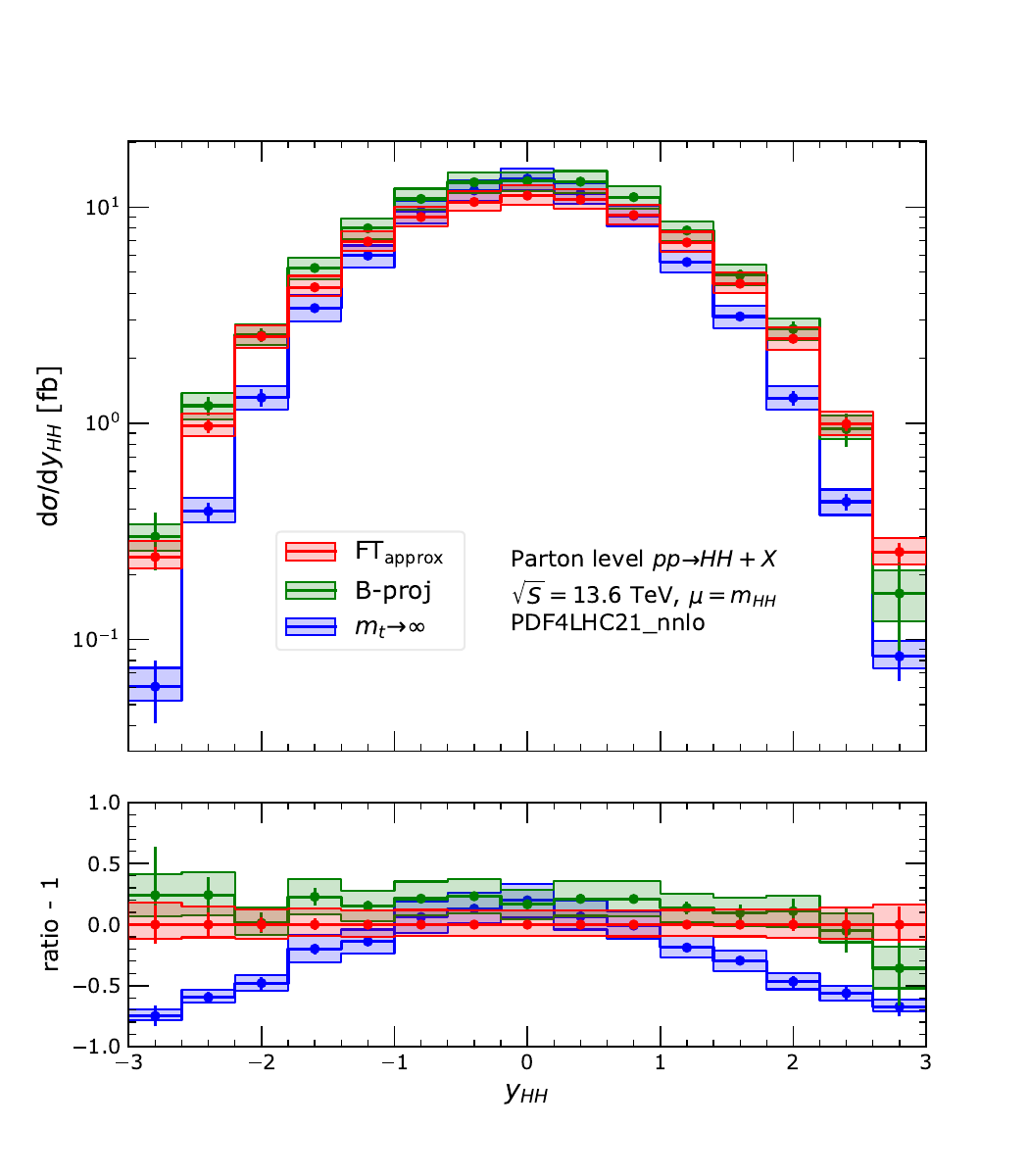} \\
  \includegraphics[width=0.49\textwidth]{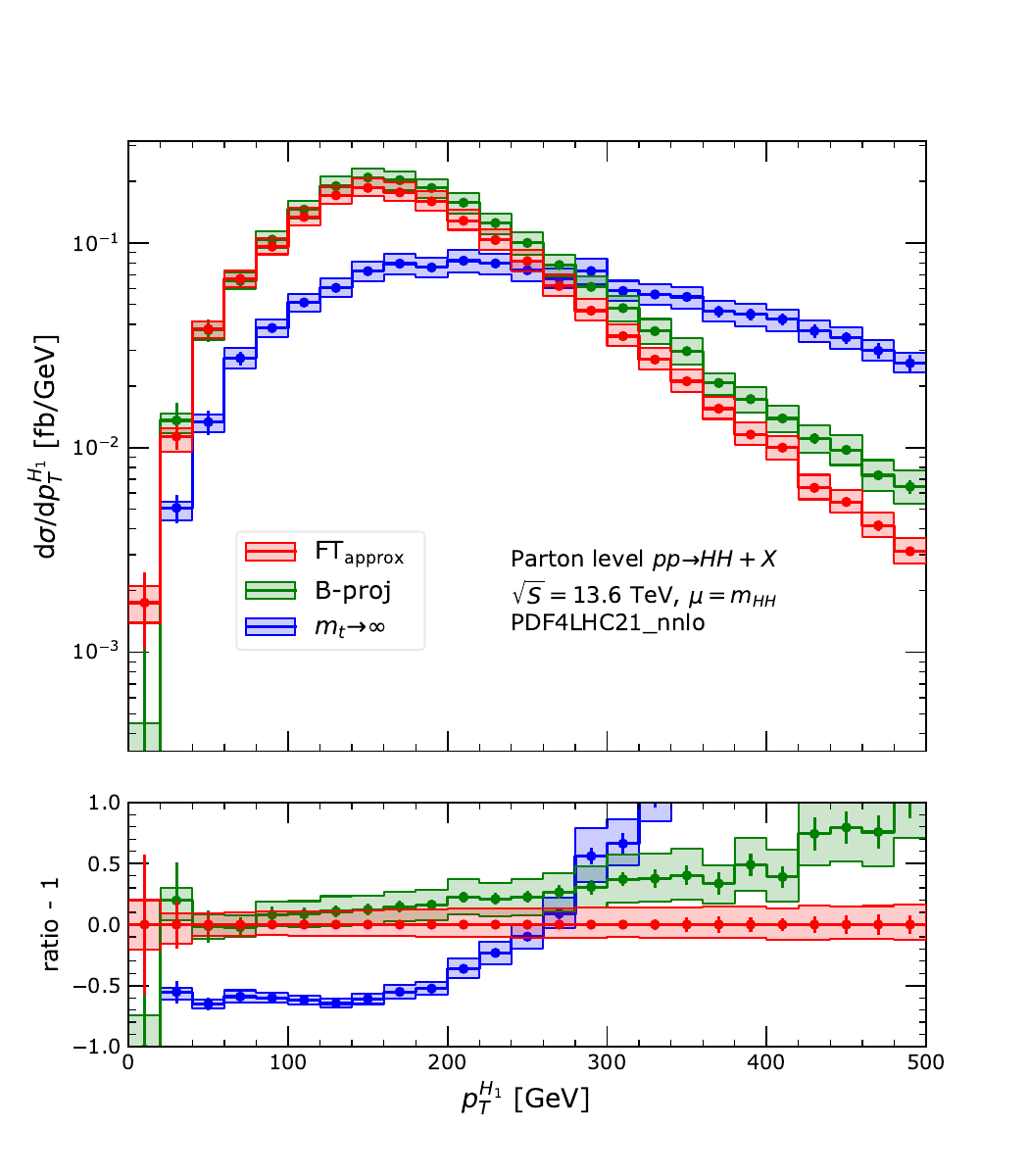} \hfill
  \includegraphics[width=0.49\textwidth]{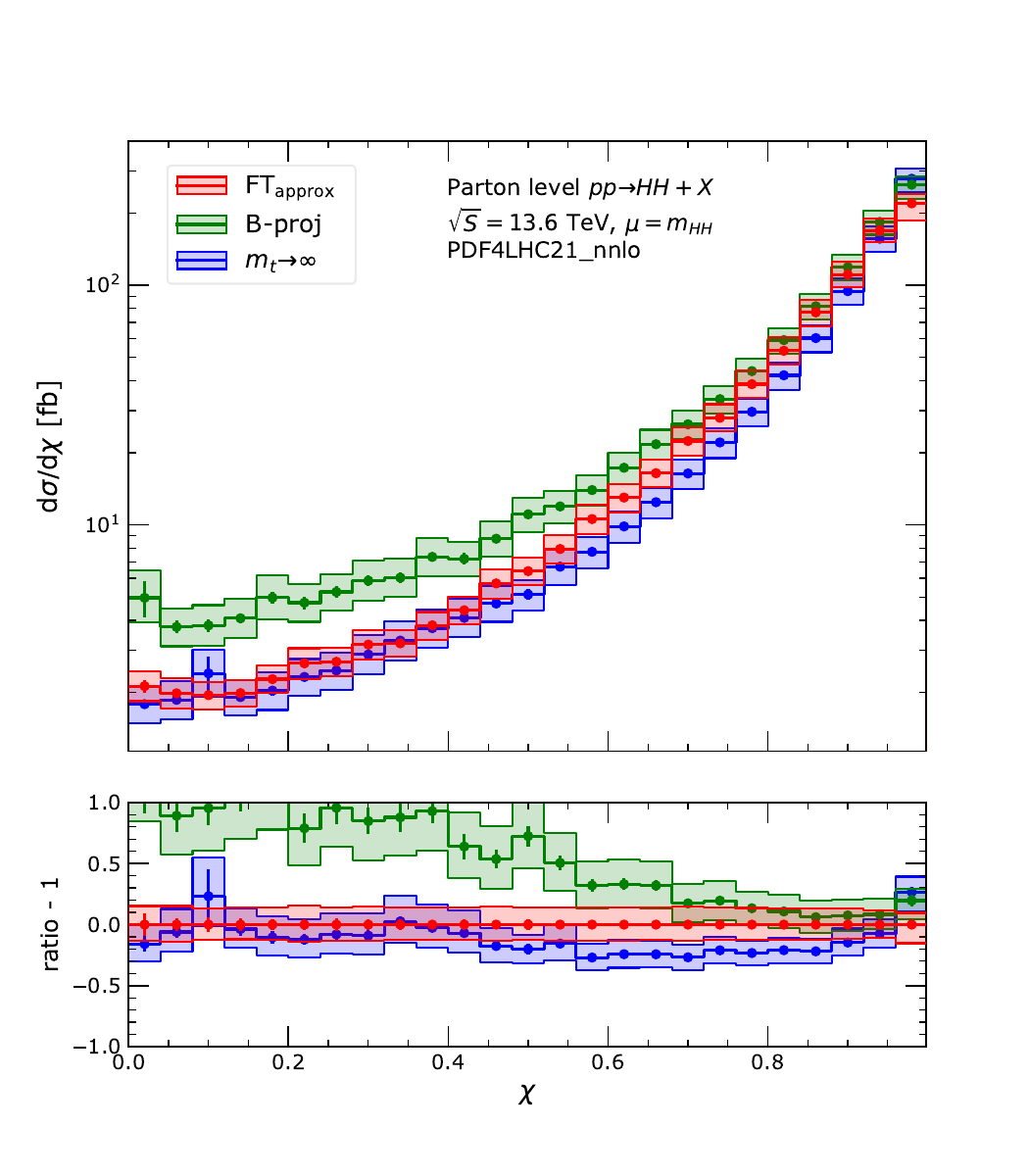}
  \caption{\label{fig:partonic-results}
  Comparison of partonic predictions obtained with the three top-quark mass approximations: $m_t \to \infty$
  (blue), B-proj (green) and \ftapprox (red), shown for various differential distributions.
  These include the invariant mass of the Higgs boson pair ($m_{HH}$, top-left),
  the rapidity of the pair ($y_{HH}$, top-right),
  the transverse momentum of the hardest Higgs boson ($p_T^{H_1}$, bottom-left)
  and the $\chi$ distribution (bottom-right).
  Results are obtained at $\sqrt{S} = 13.6$ TeV, with central scale $\mu = m_{HH}$,
  using \texttt{PDFLHC21\_nnlo} PDF set and 3-point scale variations.
  }
\end{center}
\end{figure}
% -------------------------------------------------------------------------------
Results for the three approximations are presented in
Fig.~\ref{fig:partonic-results}, where we show predictions for the invariant mass of
the Higgs boson pair $m_{HH}$, the rapidity of the pair $y_{HH}$,
the transverse momentum of the hardest Higgs boson, $p_T^{H_1}$ and
the distribution $\chi$, defined as
\begin{equation}
  \chi = \tanh \left( \frac{|y_{H_1}-y_{H_2}|}{2} \right),
\end{equation}
where $y_{H_1,H_2}$ are the rapidities of the individual Higgs
boson. For high energy Higgs bosons, this variable is directly
related to the cosine of the scattering angle of the two bosons.
Overall, we see that in most cases the B-proj prediction provides a good
approximation of the full mass effects of \ftapprox, whereas the $m_t\to\infty$
approximation fails to predict either the shape or the normalisation. This is in
agreement to what is known in the literature (see for example~\refcite{Maltoni:2014eza}).
This is particularly evident for the invariant mass distribution, where the
B-proj approximation reproduces a shape similar
to that predicted by \ftapprox, which serves as our best prediction.
However, we observe a difference in normalization between the two,
which decreases at high $m_{HH}$ but increases at
low $m_{HH}$, reaching roughly $20\%$ near the production threshold.
In contrast, the $m_t \to \infty$ approximation fails to capture both the normalization and the shape
of the distribution, and is known to provide an unreliable description for this
observable.

For the rapidity distribution $y_{HH}$, the differences between the B-proj and the \ftapprox
predictions are less pronounced. The $m_t \to \infty$ result, instead, deviates significantly in shape,
especially at large rapidities.
Overall, the impact of top-quark mass effects on this observable is modest,
especially in regions where the cross section is the largest, \emph{i.e.}~at central rapidities.
This is consistent with expectations, as this observable is mostly driven by the PDFs,
which are identical in the three approaches, and is less sensitive to the actual
matrix elements.

For the transverse momentum of the hardest Higgs boson, $p_T^{H_1}$,
the $m_t \to \infty$ approximation again significantly deviates in
both shape and normalization. The B-proj approximation also exhibits
visible discrepancies at high $p_T^{H_1}$. The
low $p_T^{H_1}$ region is particularly sensitive to resummation effects, see
Sec.~3 of \refcite{Alioli:2022dkj}. In this case, the differences
between the approximations are mostly driven by the different treatment
of the $H^{(1)}$ and $H^{(2)}$ terms in the resummation formula Eq.~\eqref{eq:convolutionmess}.
At large $p_T^{H_1}$, discrepancies arise from configurations where one Higgs boson is
very hard and recoils against a hard jet, forcing the other Higgs boson to be
soft or collinear to the beam.
In this situation, the projection mapping used in the B-proj approach in
Eq.~\eqref{eq:bprojkfact} evaluates mass corrections with a $2 \to 2$ kinematics
with back-to-back Higgs bosons, which differs significantly from the actual
$2 \to 3$ or $2 \to 4$ kinematics employed by the \ftapprox. This happens in
a region where mass corrections are large, as shown by the difference
between the \ftapprox and the $m_t\to\infty$ approximations, thus
leading to relatively large effects.

In the case of the $\chi$ distribution, the differences between
\ftapprox and $m_t \to \infty$ are less significant, indicating that top-quark
mass corrections for this observables are small.
However, in the region of small $\chi$ which is dominated by the recoil of the
Higgs boson pair system with a hard jet, B-proj predicts consistently larger values.
This discrepancy reduces only as $\chi \to 1$.
The reason for the large effect at small $\chi$ is again tied to the
same reasoning described for $p_T^{H_1}$. Indeed, in this region the correct
kinematic configurations, which would yield small mass corrections in this
observable, are those with 1 or 2 partons in the final state. However, B-proj
only accounts for mass corrections in the Born-projected kinematics, thus leading
to an overestimate of the mass effects.

In conclusion, the B-proj approximation better accounts for top-quark mass
effects than the $m_t \to \infty$ limit, in agreement with previous studies.
Nonetheless, while the reweighting procedure applied in B-proj improves the shape of some
distributions, it distorts others and leads to larger total cross
section predictions.
\ftapprox remains the most accurate prediction currently available, especially for
shape-sensitive observables such as $m_{HH}$ and $p_T^{H_1}$ (and therefore
$p_T^{HH}$), whereas for more inclusive observables like $y_{HH}$ or $\chi$, the
impact of top-quark mass corrections is comparatively mild.
%%%%%%%%%%%%%%%%%%%%%%%%%%%%%%%%%%%%%%%%%%%%%%%%%%%%%%%%%%%%%%%%%%%%%%%%%%%%%%%%

%%%%%%%%%%%%%%%%%%%%%%%%%%%%%%%%%%%%%%%%%%%%%%%%%%%%%%%%%%%%%%%%%%%%%%%%%%%%%%%%
\section{Showered results}
\label{sec:showered-results}
In this section, we present fully showered results, studying the impact of mass
corrections and their interplay with the parton shower (PS).
We begin by noting that the parton shower performs the evolution of
high-energy states down to low energy through soft and collinear emissions,
which are independent of the treatment of the mass of the top-quark in the hard
interactions. Consequently, we expect the effect of the parton shower to be
independent of the approximation chosen for top-quark mass corrections.
For this reason, while \geneva can be interfaced to other parton showers -- such
as \dire~\cite{Hoche:2015sya} and \sherpa~\cite{Sherpa:2019gpd}, as shown
in~\refcite{Alioli:2022dkj} -- we restrict our comparison to the default shower
\pythiaEight~\cite{Sjostrand:2014zea}.
In addition, since the shower performs a leading-logarithmic (LL)
resummation via its evolution, our matching procedure needs to preserve both our
higher-order accuracy in the resolution variable $\Tau_0$ and the leading
logarithmic one of any other variable performed by the shower.
The details of how \geneva interfaces to parton showers have been described
extensively in previous works, such as~\refcite{Alioli:2015toa} and
\refcite{Alioli:2022dkj}, and
remain unchanged from our earlier study in the $m_t \to \infty$
limit~\cite{Alioli:2022dkj}.
The computational setup used here is identical
to that described in section~\ref{sec:nnlo_partonic}.

\begin{figure}[H]
\begin{center}
  \includegraphics[width=0.49\textwidth]{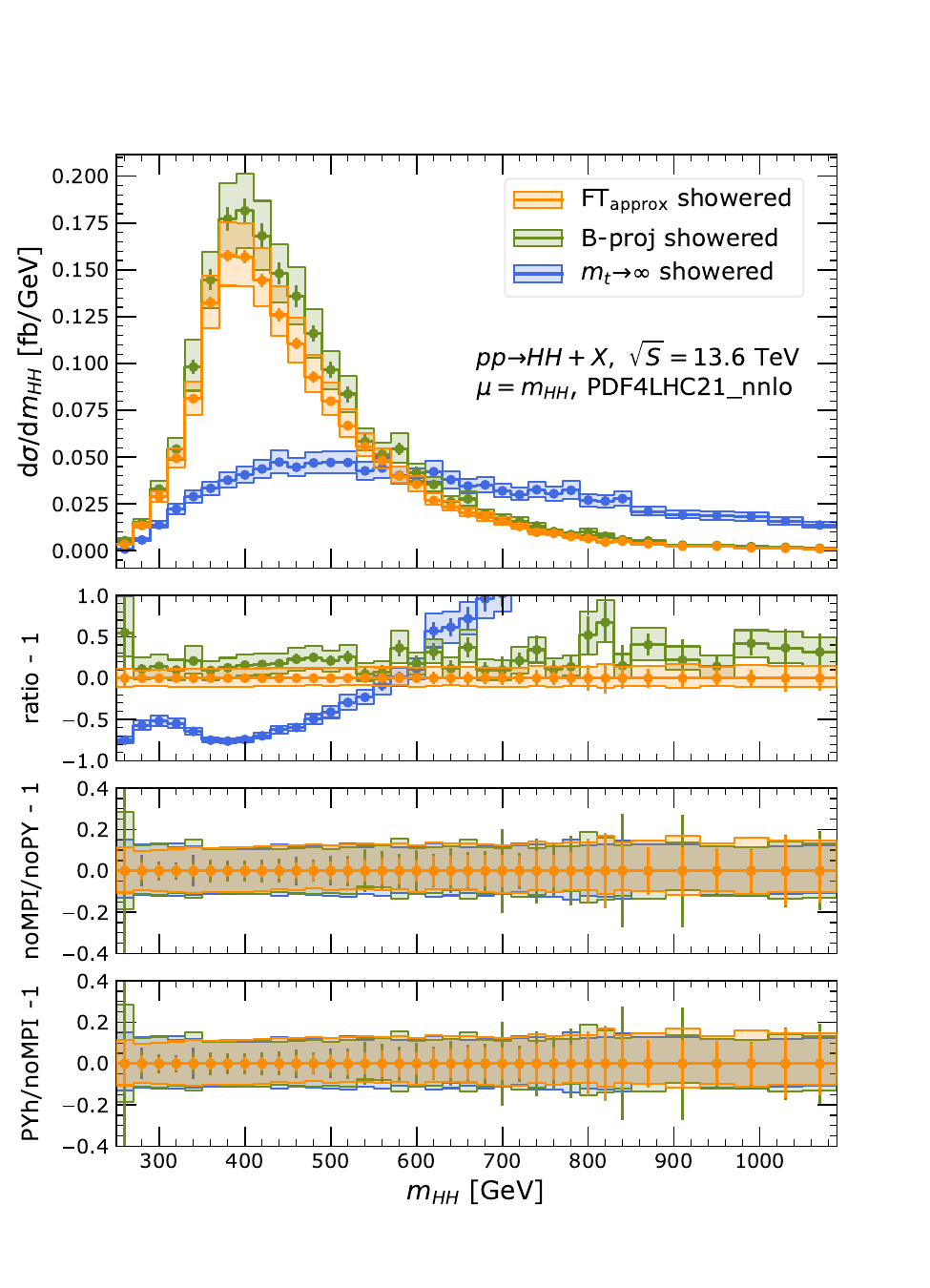} \hfill
  \includegraphics[width=0.49\textwidth]{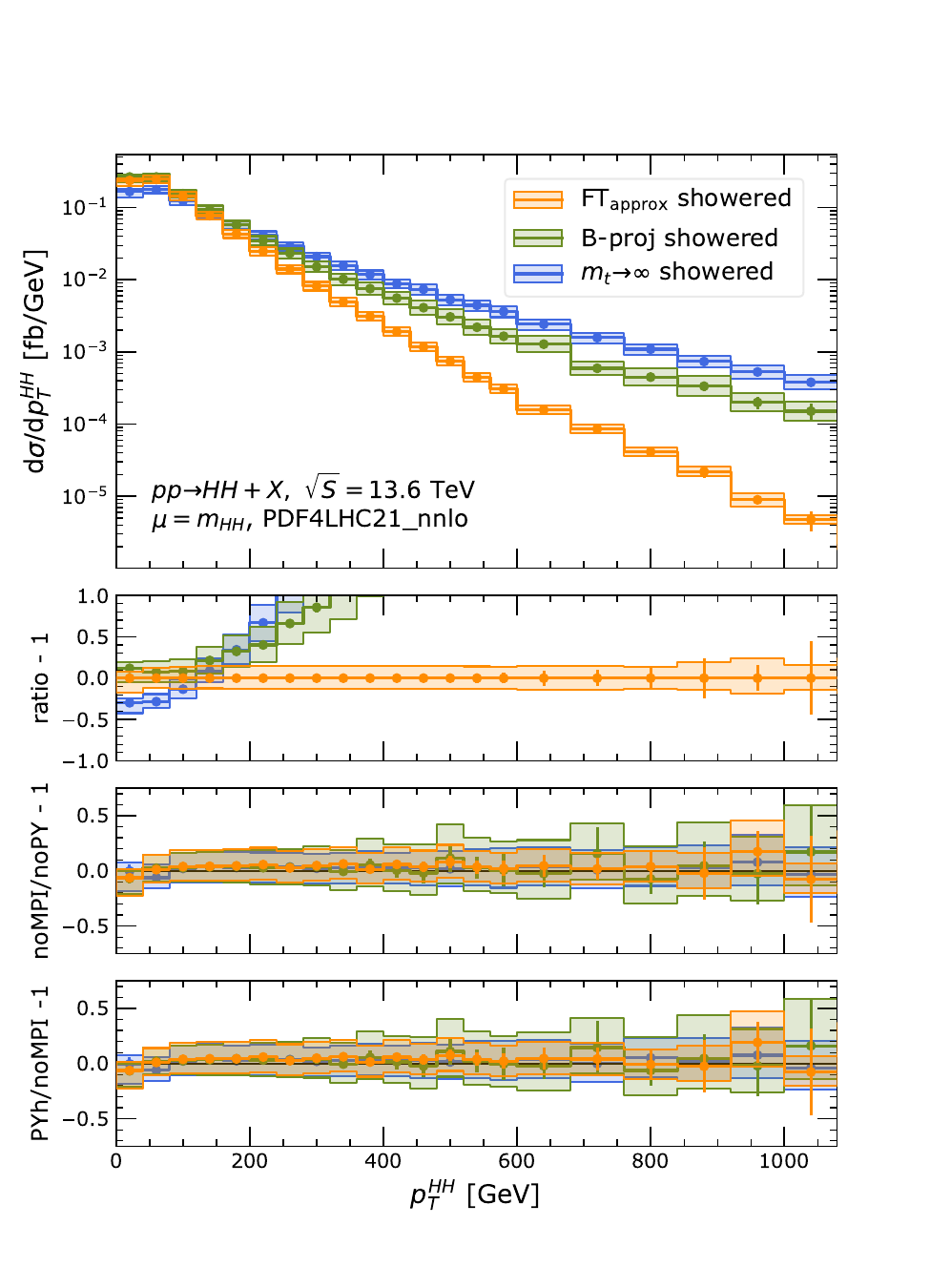}
  \\
  \vspace{-18pt}
  \includegraphics[width=0.49\textwidth]{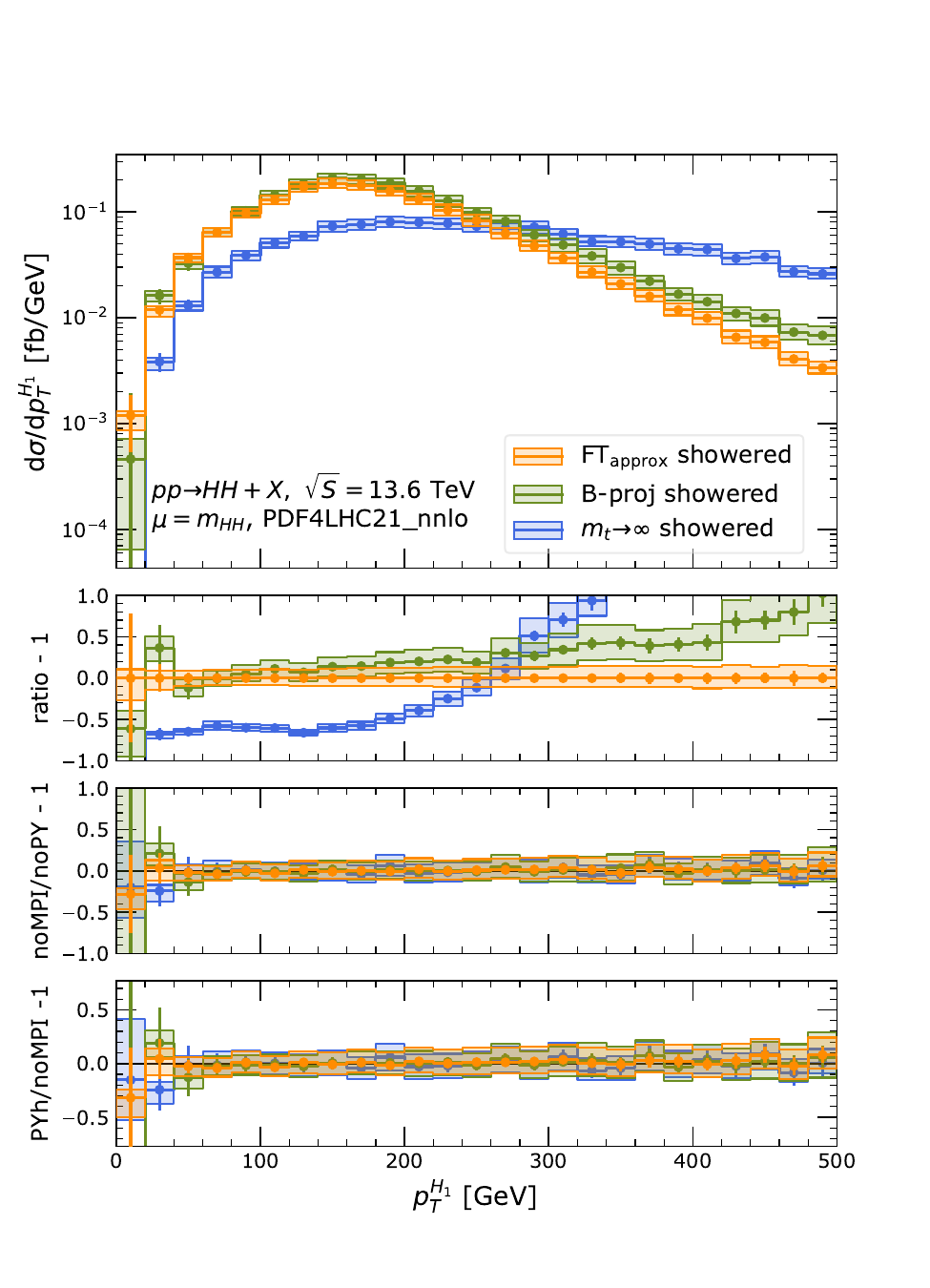} \hfill
  \includegraphics[width=0.49\textwidth]{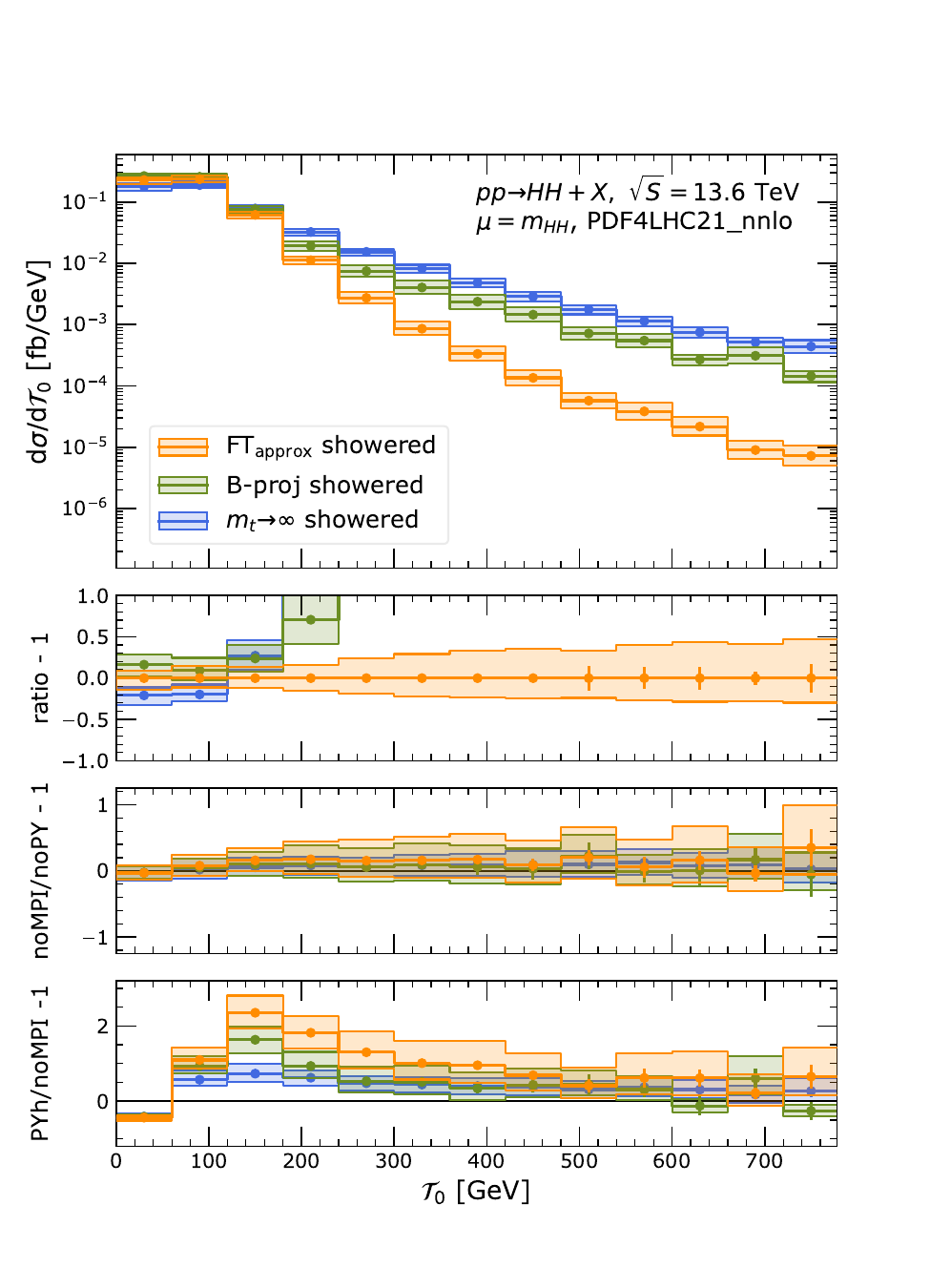}
  \caption{\label{fig:showered-results}
  \hspace{-8pt}
  Comparison of showered predictions obtained with the three top-quark mass approximations: $m_t \to \infty$
  (blue), B-proj (light green) and \ftapprox (dark orange), shown for various differential distributions.
  These include the invariant mass of the Higgs boson pair ($m_{HH}$, top-left),
  the transverse momentum of the pair system ($p_T^{HH}$, top-right),
  the transverse momentum of the hardest Higgs boson ($p_T^{H_1}$, bottom-left)
  and the $\Tau_0$ distribution (bottom-right).
  Results are obtained at $\sqrt{S} = 13.6$ TeV, with central scale $\mu = m_{HH}$,
  using \texttt{PDFLHC21\_nnlo} PDF set and 3-point scale variations.
  }
\end{center}
\end{figure}

We present our results in Fig.~\ref{fig:showered-results} for the three
top-quark mass approximations, across four differential distributions: the
invariant mass of the Higgs boson pair $m_{HH}$,  their transverse momentum
$p_T^{HH}$, the transverse momentum of the hardest Higgs boson, $p_T^{H_1}$ and
finally the zero-jettiness $\Tau_0$.
The structure of the results is the following. The main panel displays
predictions after the full shower procedure, including hadronisation
and multi-parton interaction (MPI) effects, but excluding hadron decays.
The first ratio panel shows the relative difference of each
approximations with respect to the \ftapprox result. The
second ratio panel illustrates the impact of the parton shower at parton level (i.e,
without hadronisation and MPI effects), by showing the relative difference
with respect to the partonic results as described in
Sec.~\ref{sec:nnlo_partonic}.
The last ratio panel displays the impact of hadronisation and MPI
effects relative to the parton-level shower results (no hadronisation
and no MPI).

We start by noting that in the case of the Higgs boson pair mass and the $p_T$ of the
hardest Higgs boson, the inclusion of top-quark mass effects produces corrections
of similar size at hadron level as at parton level, as presented in
Sec.~\ref{sec:nnlo_partonic}. This can be seen by comparing the first ratio
panel of Fig.~\ref{fig:showered-results}  with the ratio panel of
Fig.~\ref{fig:partonic-results} for these observables. This further validates our assumption
that the shower behaves independently of the treatment of top-quark
mass effects in the hard interaction.
In the second panel of each plot, we report the relative impact of the parton
shower without hadronisation and MPI effects to the same prediction at parton
level. The last panel then shows the relative impact of hadronization and
MPI effects relative to the parton shower of the previous panel.
Since the invariant mass of the Higgs boson pair system is an inclusive
observable, for which NNLO accuracy is achieved, the parton shower has no effect due
to the unitarity of the shower matching.
This also holds true, in particular, for
hadronisation and MPI effects, as seen in the third panel.
%------------------------------------------------------------------------------%
% ------------------------------------------------------------------------------%
For the transverse momentum distributions, $p_T^{HH}$ and $p_T^{H1}$,
the parton shower modifies only the first few bins, with effects
of approximately $5\%$ relative to the \geneva partonic predictions.
These effects are consistent across the three approximations
and are concentrated in the low-$p_T$ region, where
resummation effects dominate and are captured by the shower.
Hadronisation and MPI effects, shown in the third panel, are of similar size to the
pure shower effects seen in the second panel.
As a consequence, they are of moderate size and concentrated in the low-$p_T$
region.
Turning to the $\Tau_0$ distribution, in the \geneva framework
the matching between the partonic events and the parton shower ensures that
the accuracy of each observable is, at least, as accurate as the parton shower
is, which in this case is LL accurate. At the same time,
we enforce that the accuracy of the resolution variable for which we perform the
resummation is also not spoiled by the parton shower. The result is that any
numerical modifications introduced by the shower in the small $\Tau_0$ region
should not exceed the typical size of higher-order resummation effects (see
\emph{e.g.} Fig.~11 of~\refcite{Alioli:2022dkj}).
Indeed, as seen in the second panel plot, the shower does not significantly alter the
distribution within statistical uncertainties. This stability is partly due to
the relatively wide binning used for this observable. Lastly, we see that the
impact of hadronisation and MPI for $\Tau_0$ are instead large, as shown in the
third panel. This agrees with previous studies which showed the sensitivity of
the zero-jettiness to such effects~\cite{Alioli:2016wqt}.
In particular, we see that their influence is mostly visible at low values of
$\Tau_0$, where it significantly distorts the spectrum, gradually diminishing
toward the tail of the distribution.

In conclusion, the qualitative impact of the default parton shower remains
consistent with the results of our previous study using the $m_t \to
\infty$ limit. Similar patterns are observed across all
top-quark mass approximations considered in this work.
%%%%%%%%%%%%%%%%%%%%%%%%%%%%%%%%%%%%%%%%%%%%%%%%%%%%%%%%%%%%%%%%%%%%%%%%%%%%%%%%

%%%%%%%%%%%%%%%%%%%%%%%%%%%%%%%%%%%%%%%%%%%%%%%%%%%%%%%%%%%%%%%%%%%%%%%%%%%%%%%%
\section{Conclusions}
\label{sec:conclusions}
The process $gg \to HH$ plays a central role in the exploration of the Higgs sector,
particularly in probing the Higgs boson self-interaction via the trilinear coupling $\lambda_{HHH}$.
Given its small cross section and the complexity of the final state, precise and
reliable theoretical predictions are essential not only for improving signal
sensitivity, but also for disentangling potential deviations from Standard Model
expectations.
In this work, we have extended our previous implementation of double Higgs boson
production via gluon fusion in the \geneva framework, which was originally performed in
the $m_t \to \infty$ limit, by including different approximations that account
for finite top-quark mass effects.
Specifically, we have added support for the so-called Born-improved (B-proj)
and the \ftapprox approximations.
The former approach reweights each event by the ratio of the full-theory (massive)
to $m_t\to\infty$ Born-level squared amplitudes, requiring a projection onto
Born-like kinematics.
The \ftapprox method instead incorporates all known ingredients with full top-quark
mass dependence with the correct kinematics.
The remaining unknown components -- namely the double-virtual and real-virtual corrections --
are rescaled using Born-level matrix elements computed with exact top-quark mass dependence.

By leveraging the \geneva framework, we are able to produce fully differential
NNLO predictions matched to a  $\Tau_0$ resummation at NNLL$^{\prime}$ accuracy
within the SCET formalism, and further matched to a parton shower including
hadronisation and multi-parton interactions (MPI). We have validated our
partonic predictions against standard fixed-order results obtained with \Matrix,
and we have studied the impact of the parton shower, hadronisation, and MPI on
key differential observables.

This constitutes the first NNLO+PS implementation of double Higgs boson
production via gluon fusion that includes all currently known
top-quark mass corrections in a realistic and flexible event
generator, ready to be used in experimental analyses at the LHC and
future hadron colliders. To provide a framework capable of estimating all
sources of theoretical uncertainties, we plan to include in future versions of
our code various improvements. These will comprise top-quark mass variations, EW
corrections, at least in approximate form, as well as bottom-quark mass effects,
as soon as they become available.
%%%%%%%%%%%%%%%%%%%%%%%%%%%%%%%%%%%%%%%%%%%%%%%%%%%%%%%%%%%%%%%%%%%%%%%%%%%%%%%%

%%%%%%%%%%%%%%%%%%%%%%%%%%%%%%%%%%%%%%%%%%%%%%%%%%%%%%%%%%%%%%%%%%%%%%%%%%%%%%%%
\section*{Acknowledgments}
\label{sec:Acknowledgements}
We thank our \geneva collaborators for their work on the code and in particular A.~Broggio for discussions.
We also thank J.~Mazzitelli for providing the values for the \Matrix results
appearing in \refcite{Grazzini:2018bsd}.
This project acknowledges funding from the European Research Council (ERC) under
the European Union’s Horizon 2020 research and innovation program
(Grant agreements No. 714788 REINVENT and 101002090 COLORFREE) and also
acknowledges support from the Deutsche Forschungsgemeinschaft (DFG) under
Germany’s Excellence Strategy – EXC 2121 “Quantum Universe”– 390833306. We acknowledge
financial support, super-computing resources and support from ICSC – Centro Nazionale
di Ricerca in High Performance Computing, Big Data and Quantum Computing – and
hosting entity, funded by European Union – NextGenerationEU.
%%%%%%%%%%%%%%%%%%%%%%%%%%%%%%%%%%%%%%%%%%%%%%%%%%%%%%%%%%%%%%%%%%%%%%%%%%%%%%%%

%%%%%%%%%%%%%%%%%%%%%%%%%%%%%%%%%%%%%%%%%%%%%%%%%%%%%%%%%%%%%%%%%%%%%%%%%%%%%%%%
%% Bibliography
\bibliographystyle{JHEP}
\bibliography{geneva}

\end{document}